\documentclass[useAMS,usenatbib]{mn2e}
\usepackage{graphicx}
\usepackage{amsmath}

\DeclareGraphicsExtensions{.pdf, .jpg, .png}

\graphicspath{{}}

\title[Stirring N-body systems: Universality of end states]{Stirring N-body systems: Universality of end states}
\author[Jeremy A. Barber, Hongsheng Zhao, Xufen Wu and Steen H. Hansen]{Jeremy A. Barber$^{1}$\thanks{E-mail:
jab22@st-andrews.ac.uk (JAB); hz4@st-andrews.ac.uk (HZ); xwu@mpe.mpg.de (XW); hansen@dark-cosmology.dk (SH)}, Hongsheng Zhao$^{1}$, Xufen Wu$^{2}$ and Steen H. Hansen$^{3}$\\
$^{1}$Department of Physics \& Astrophysics, University of St Andrews, North Haugh, St Andrews, Fife, KY16 9SS, UK\\
$^{2}$Max-Planck-Institut fur Extrasterrestrische Physik, Postfach 1312, D-85741 Garching, Germany\\
$^{3}$Dark Cosmology Centre, Niels Bohr Institute, University of Copenhagen, Juliane Maries Vej 30, 2100 Copenhagen, Denmark}
\begin{document}

\date{Accepted ----. Received ----}

\pagerange{\pageref{firstpage}--\pageref{lastpage}} \pubyear{2011}

\maketitle

\label{firstpage}

\begin{abstract}
We study the evolution of the phase-space of collisionless N-body systems under repeated stirrings or perturbations. We find convergence towards a limited solution group, in accordance with \citet{Hansen2010}, that is independent of the initial system and environmental conditions, paying particular attention to the assumed gravitational paradigm (Newtonian and MOND). We examine the effects of changes to the perturbation scheme and in doing so identify a large group of perturbations featuring radial orbit instability (ROI) which always lead to convergence. The attractor is thus found to be a robust and reproducible effect under a variety of circumstances.
\end{abstract}

\begin{keywords}
galaxies: haloes, galaxies: kinematics and dynamics, methods: N-body simulations, methods: numerical
\end{keywords}

\section{Introduction}
One of the ongoing problems in the field of galactic dynamics is the non-keplerian nature of rotation curves in spiral galaxies \citep{Salucci2007} and, consequently, the inferred presence of massive but undetectable structures that envelope the luminous component of galaxies. According to the prevailing cosmological models, this structure is comprised of weakly-interacting particles which are known as Dark Matter (DM). Of particular interest in the study of DM is the characteristic density profiles of the halos that surround observed baryonic structures as, without insight into the forms taken by these halos, it is hard to make meaningful predictions in the theory. Cosmological N-body simulations suggested early on that all stable halos should look like isothermal spheres that could be fit by some universal profile \citep[and references therein]{Dubinski1991}. That profile, named the `NFW profile' after \citet{Navarro1996}, was found to be a two-power model with parameters $\alpha=1, \beta=3$ as follows:

\begin{equation}
\label{eqn:nfw}
\rho(r)=\frac{\rho_0}{\left(\frac{r}{a}\right)(1+\frac{r}{a})^{2}}
\end{equation}

Due to both the simplicity of the model and the apparent universality of the result, NFW profiles have become accepted as the `go-to' model for simulating halo characteristics with many results being based on them. Because of this, it is vital that the ubiquity of the NFW profile be properly understood and the fact that it is not is a cause for concern. This is especially true since the universality of the profile has been called into question in the past by X-ray observations \citep{Makino1998} and the Tully-Fisher relation \citep{McGaugh1998}.

One potential problem is that in the original study, \citet{Navarro1996}, only fully-equilibriated halos were selected from the initial, low-resolution simulation for further, high-resolution simulation. Subsequent investigations into the larger population have shown that while the profile is still a decent fit, even for non-equilibrium structures \citep{Jing2000}, there is evidence that a less global profiling system may have to be used instead \citep{Host2011}.

We focus on recent work by \citet{Hansen2010} (hereafter HJS) where it was suggested that all collisionless systems will tend to move towards characteristics drawn from narrow range if they are gently perturbed from their current equilibrium and are allowed to find a new one. HJS used a simple algorithm to disturb a set of relaxed systems multiple times and observed, in each, a tendency for each successive equilibrium to converge to a particular region in the allowed parameter space. We attempt to test the reproducibility of this phenomena, examine how easily the behaviour can be disturbed and attempt to quantify and explain this behaviour. We do not focus on other cosmological simulation work in this paper as our aim here is to reproduce and understand the mechanism behind the attractor. It is not the aim of this paper to discuss the appearance or otherwise of the attractor in either observation or simulation as it is premature to discuss the physical relevance of an effect that is not understood. Future work will use this paper as a foundation to discuss the physical relevance of the attractor after the mechanism and phenomenology are fully understood.

\section[]{Attractors in the Jeans' parameter space}
\begin{figure}
\includegraphics[width=84mm]{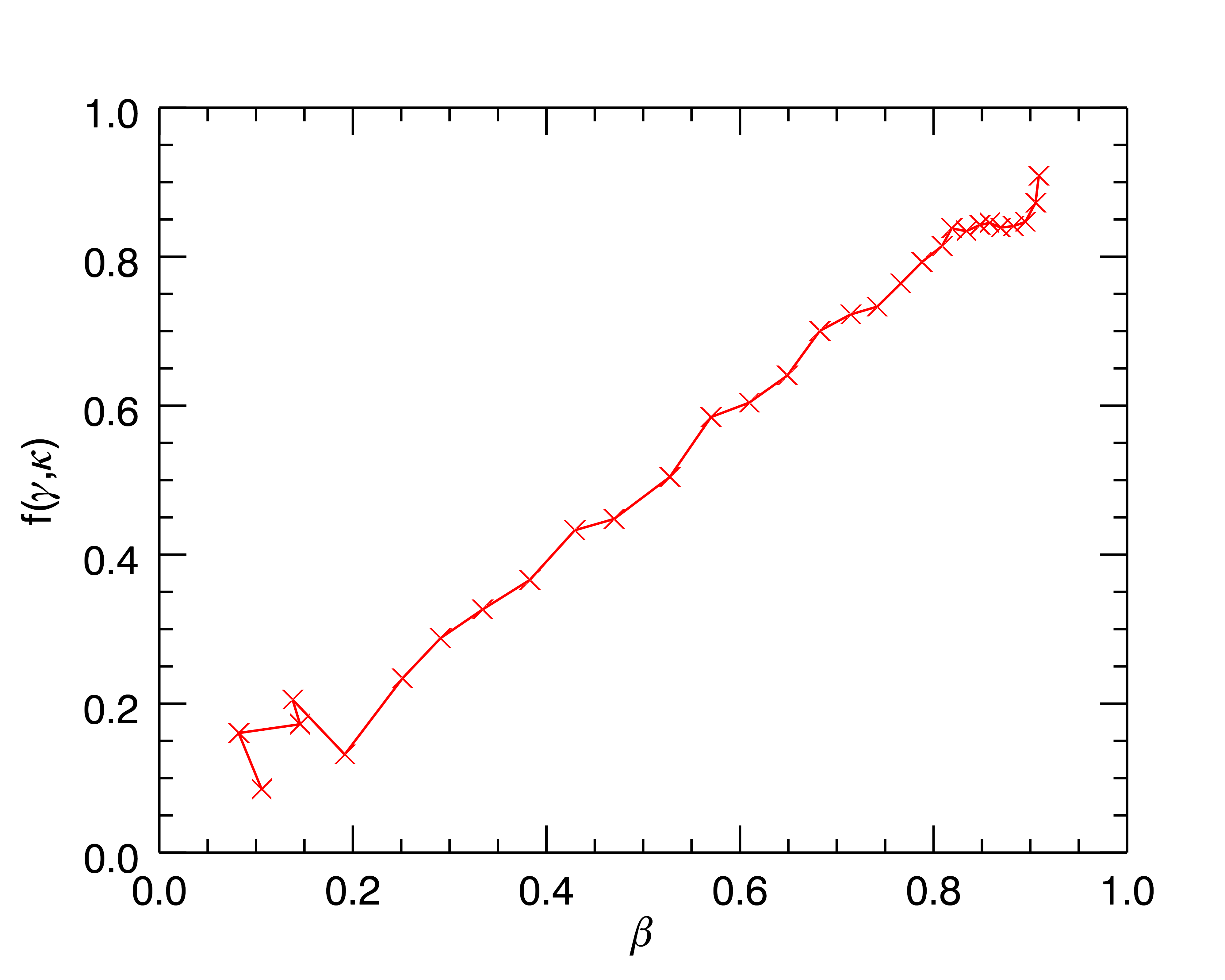}
\caption{\label{fig:onedim} 1-D form of the attractor in the parameter space plotting HJS's data for $\beta$ against the $f(\gamma,\kappa)$ given in equation \ref{eqn:attractor}.}
\end{figure}

The Jeans equations \citep{Jeans1915} describe the relationship between a density field, $\nu(x,t)$, a potential, $\Phi(x,t)$ and the arrangement of velocity vectors in the system. In most cases there is not enough information to find a single, unique solution for an unknown component and one must accept a range of permitted solutions instead. For a system with no net velocity, such as the ones used in this work, a useful form of the equation is the following:

\begin{equation}
\label{eqn:jeans}
v_c^2=-\sigma_r^2(\gamma+\kappa+2\beta)
\end{equation}

given $\sigma_r^2$ is the radial velocity dispersion, $\sigma_t^2$ is the tangential velocity dispersion, $v_c^2$ is the circular speed and the other terms are as follows:

\begin{equation}
\label{eqn:definitions}
v_c^2=\frac{GM_{tot}}{r}; \gamma=\frac{d\log{\rho}}{d\log{r}}; \kappa=\frac{d\log{\sigma_r^2}}{d\log{r}}; \beta=1-\frac{\sigma_t^2}{\sigma_r^2}
\end{equation}

where $M_{tot}$ is the total mass of the system. There are limitless configurations for which this equation will hold. It is this that makes any convergent result so surprising; that in an effectively infinite parameter space only one subset of results should be favoured. The result that HJS found was a strong link in the parameter space between the quantities $\beta$ and $\gamma+\kappa$ to which they fit an empirical relationship thus:

\begin{equation}
\label{eqn:attractor}
\beta=\frac{-0.15\gamma-0.85\kappa}{(1+(-0.15\gamma-0.85\kappa)^3)^{\frac{1}{3}}}
\end{equation}

which removes a degree of freedom from the system. HJS find that equation \ref{eqn:attractor} is an `attractor' i.e. a solution that systems will converge towards if they are free to move in it's parameter space. In order to observe evolution towards the attractor we need to perturb an initially equilibriated system in some physical meaningful way and see if it settles into equilibria successively closer to one particular solution. The attractor can be represented as a linear relation when projected into the right parameter space such as in figure \ref{fig:onedim}. The data in this plot is taken from the final states of the simulations in HJS and it is that data which will provide us with our attractor curve throughout this work.

However, since this space is somewhat unintuitive, we shall usually plot the attractor in the more accessible parameter of space $\beta$ versus $\gamma+\kappa$. In this space the attractor has a more complex 'S-bend' shape which shall be seen in subsequent plots.

\section[]{Numerical simulations}

\subsection{Initial conditions and numerical code}
The process of perturbation and relaxation is carried out using N-body methods designed to replicate those used by HJS. Initial conditions are generated using the method outlined in \citet{Gerhard1991}. This approach is based on splitting the distribution function of the system, $f(\epsilon,L)$, into two separate functions: an energy distribution function, $g(\epsilon)$, which controls how energy levels are populated and a circularity function, $h(x)\equiv h(\frac{L}{L_0+L_{circ}(\epsilon)})$, which controls how circular the orbits of energy $\epsilon$ are. From these two functions one can directly produce functions for density and velocity dispersion necessary for building a Jeans'-stable system, for example:

\begin{equation}
\rho(r)=\frac{4\pi}{r^2}\int^{\psi(r)}_0d\epsilon\,g(\epsilon)[L_0+L_c(\epsilon)]\left[\int^{x_0(\epsilon)}_0\frac{dx\,xh(x)}{(x^2_0(\epsilon,r,L_0)-x^2)^{\frac{1}{2}}}\right]
\end{equation}

with comparable formulae for other velocity dispersions. Accordingly, a desired profile can be analytically generated by the appropriate choice $h(x)$ (with $g(\epsilon)$ being subsequently derived using Lucy's method) and that profile can be randomly populated by sampling the $f(\epsilon,L)$.

Our initial systems are Plummer spheres consisting of 750,000 particles with a total mass of $5\times 10^8$ $M_{\sun}$. We chose to only investigate one density profile in detail as it allows direct comparison of the effects of different perturbation schemes. Plummer spheres were chosen as they are formally unrelated to the NFW profile and are easy to create with varying anisotropies. We create a variety of Plummer spheres to test various aspects of the attractor. We also use one set of initial conditions designed to be, at any radius, more strongly anisotropic than the attractor. This system has a power law density gradient of $r^{-2}$ throughout and a strongly radially anisotropic profile.

We make use of the N-body code NMODY, a particle mesh code capable of implementing MOND, for our numerical simulation. NMODY uses an iterative scheme alongside a standard particle mesh to compute Newtonian forces and implements an extension to these for working in MOND as follows. Firstly, the code examines the density distribution and approximates it as an exact spherical solution, or takes the previous state of the system, as the starting point. The difference between the assumed density distribution and the actual distribution is assessed by assessing the convergence between the acceleration field produced by the density and the density implied by the potential. When the two are, to within some user-defined accuracy, producing the same accelerations then that potential is used. The code uses a standard leapfrog scheme to apply timesteps. For a full description of the code see \citet{Ciotti2006}.

The ability to have simulations run in MOND is used to examine whether using a different form of Poisson's equation has any impact on the attractor. If the attractor is a gravitational effect then it would be expected that it would be fundamentally linked to the description of gravity being used. Thus, we can test whether this is true by altering the fundamental equations governing gravity in our simulation and attempting to identify any impact that this has on the attractor. Since MOND performs just this kind of fundamental alteration to gravity a simulation run in the MOND paradigm should reveal whether the attractor arises from gravity or not depending on whether or not the convergence is different. It is because of this that MOND is useful in narrowing down the mechanism that drives convergence.

Since MOND is used primarily as a tool for altering Poisson's equation we will not deal in depth with the theory itself or the mathematics behind it. Suffice to say here that MOND strengthens gravity compared to Newtonian gravity but only in regimes of very low acceleration. Thus we have our three main paradigms of Newtonian gravity, `perturbative' or `weak' MOND where the accelerations are high and the MONDian modifications are only a low order perturbation and `deep' or `strong' MOND where the bulk of the system is in the low acceleration regime and the MOND effect is significant. In order to maintain the same density profile and total mass between all three scenarios while still operating in the relevant acceleration regimes, the scale radii of the models are bigger for models with a stronger MOND influence: Newtonian simulations use 0.05$\,$kpc, weak MOND also uses 0.05$\,$kpc and deep MOND 1.0$\,$kpc. For more details please see Appendix \ref{sec:MOND}, \citet{Bekenstein1984}, \citet{NMODY} or the introduction of \citet[and references therein]{Ciotti2006}.

\subsection[]{Basic Perturbation}
After choosing which model to use for our initial conditions (ICs) we define a simple algorithm that we hope will give us our evolution towards an attractor. Our principle algorithm is taken from that used by HJS but with some minor differences such as having fixed-mass bins rather than fixed-radius bins and applying different factors to each velocity component rather than the same one three times. The binning in the simulation allows us to define conservation laws for localised groups of particles in the simulations rather than only conserving over the simulation as a whole, softening the impact of the perturbation. By contrast, the binning in the analysis is construct quantities such as velocity dispersion and anisotropy which break down for single particles. For the former we choose to define bins as radial shells containing a fixed number of particles as it enables good statistics for the conservation laws in the outer edges of the system where number densities are lower. Other schemes are employed to test various components of the attractors behaviour and they will explained in their respective sections.

\begin{itemize}
\item Set up a series of radial bins. We chose to create bins defined to contain 5,000 particles.
\item For each particle in each bin we examine each of the three orthogonal velocity vectors and multiply each by a random number drawn from a uniform distribution centered around unity e.g. $1-0.25<f<1+0.25$. This is referred to as the `shock' or `perturbation' and $f$ can be called the `kick scale factor'. As a shorthand $f$ is expressed as $\pm n$, so the previous example would be communicated as $f=\pm0.25$.
\item Make a choice about what quantities to conserve in the system. Here we shall only deal with energy conservation as we currently have no algorithm that can conserve both energy and angular momentum without either failing to converge or introducing biases. For an example of an equivalent procedure for angular momentum please refer to Appendix \ref{sec:append}
\item To conserve energy, the energy in the bin is reassessed in order to rescale all the velocity components equally to enforce conservation. Note that, since we are not moving particles around, only the kinetic components before and after, $T_{init}$ and $T_{final}$, need consideration:

\begin{equation}
\label{eqn:econ}
v_{i,j,k}=v_{i,j,k}\sqrt{\frac{T_{init}}{T_{final}}}
\end{equation}

\item If a particle has been perturbed such that it is no longer bound in the system then it is re-randomised and the conservation algorithm is reapplied.
\item Derive a dynamical timescale for the system

\begin{equation}
\label{eqn:time}
t_{dyn}=\sqrt{\frac{1}{G\rho}}\text{ where }\rho=\frac{0.95\times M_{tot}}{\frac{4}{3}\pi{r_{95\%}^3}}
\end{equation}

where we are using the $95^{th}$ mass percentile as a representative distance for the system. For our initial systems this is equivalent to approximately 3 scale radii.

\item The system is then left to evolve in an N-body simulator for 1 dynamical timescales. This `flow' period allows the system to relax and find a new equilibrium. If we were to apply another shock too soon then the impact of the second shock would be indistinguishable from that of the first.
\item The entire `shock-flow' cycle is repeated 30 times.
\end{itemize}

\subsection{Analysis}
Where possible, information about the system is taken directly from the `per-particle' position and velocity file or the diagnostic output from the N-body simulation. Local density and velocity dispersion are interpolated onto a spherical polar grid. Particles have a linear smoothing kernel applied to them and the mass contribution at a point of interest are summed to find the local density. In the case of the velocity dispersion we apply the same method but weight the particles by their velocities while applying the smoothing kernel. This smooths the small scale noise that otherwise tends to dominate the dispersion. We choose points arranged on a spherical grid and average over the angular space to find radial profiles. Note that the use of a spherical polar grid here means that our analysis uses bins of radius in contrast to our simulations which use bins of equal mass.

\section[]{Recreating the HJS results}
\subsection{Outline of simulations}
Various simulations are run over the course testing many aspects of the attractor. Table \ref{tab:table} summarises, for clarity, what has been run and the order in which it will be discussed.

\begin{table*}
 \centering
 \label{tab:table}
 \begin{minipage}{140mm}
  \caption{Outline of all simulations performed.}
  \begin{tabular}{@{}lrrr@{}}
  \hline
   Gravity & IC $\beta$ profile & Kick scale factor & Flow time ($T_{dyn}$) \\
 \hline
 Newtonian & Isotropic & $\pm 0.5$ & 1 \\
 Newtonian & Isotropic & $\pm 0.5$ & 3 \\
 Newtonian & Isotropic & $\pm 1.0$ & 1 \\
 Newtonian & Isotropic & $\pm 1.0$ & 3 \\
 Newtonian & Radial & $\pm 0.5$ & 3 \\
 Newtonian & Tangential & $\pm 0.5$ & 3 \\
 Newtonian & Isotropic & $\pm 0.5$ (spherical coordinates) & 3 \\
 Newtonian & Above attractor & $\pm 0.5$ & 3 \\
 MONDian & Isotropic & $\pm 0.5$ & 3 \\
 MONDian & Radial & $\pm 0.5$ & 3 \\
 MONDian & Tangential & $\pm 0.5$ & 3 \\
 Deep MONDian & Isotropic & $\pm 0.5$ & 3 \\
 Newtonian & Isotropic & bimodal & 3 \\
 Newtonian & Isotropic & Attempt to conserve L and E & 1 \\

\hline
\end{tabular}
\end{minipage}
\end{table*}

Terms such as `bimodal kick' are explained in detail in their relevant sections.

\subsection{The assumption of spherical symmetry}
\label{sec:sphersym}
\begin{figure}
\includegraphics[width=84mm]{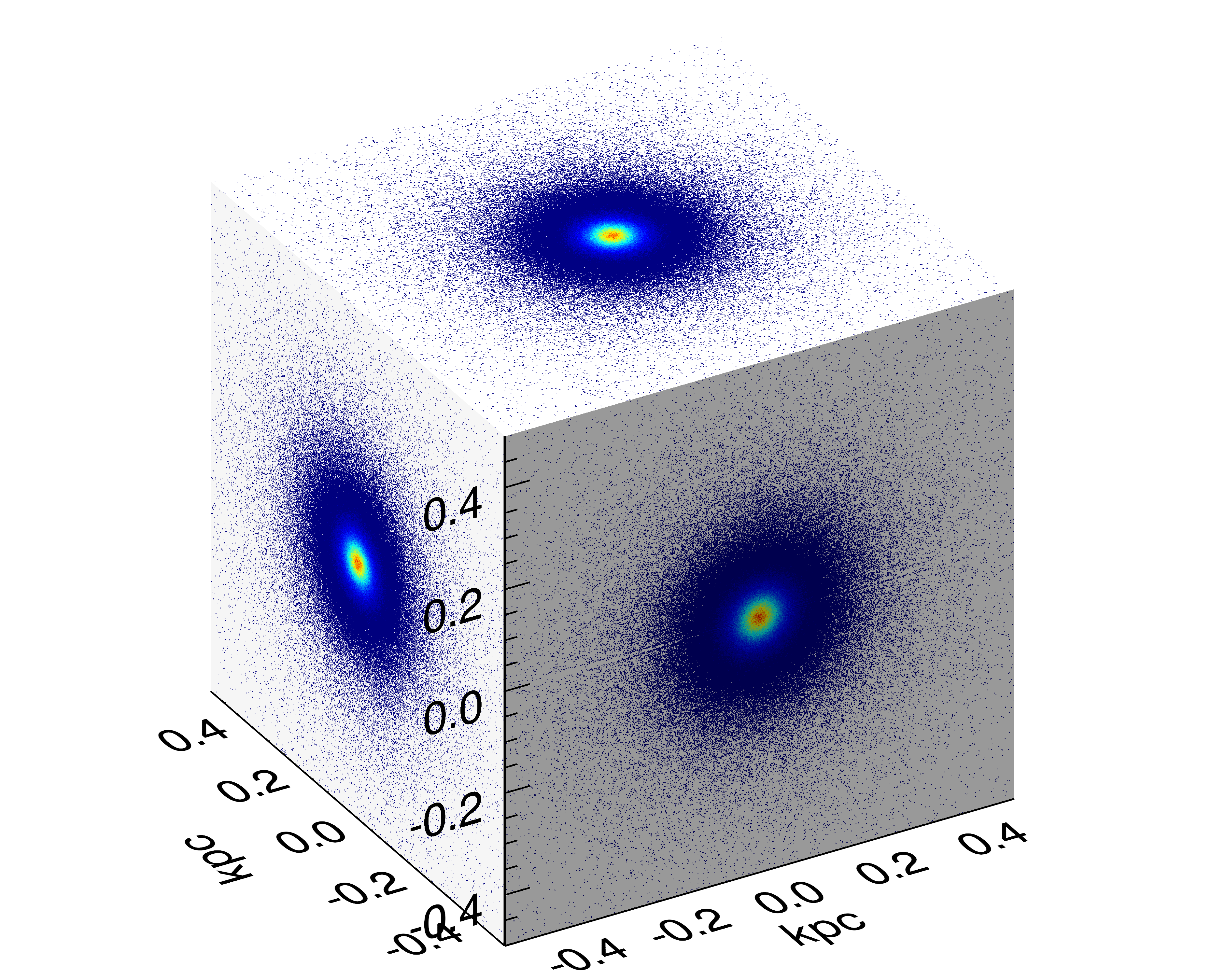}
\caption{\label{fig:spacesymmetry}Model particle densities projected along orthogonal lines of sight into the system.}
\end{figure}

Before any in depth analysis can be made it is important to learn whether it is reasonable to compress the available phase-space, $f(r,\theta,\phi,v_r,v_{\theta},v_{\phi})$, according to spherical symmetry, $f(r,v_r,v_t)$, as we would like to assume from the symmetries of both our IC's and our perturbation algorithm. This is important as failure to account for lack of sphericity could cause the system to appear more isotropised than before (see Appendix \ref{sec:append2}). First we must prove that our perturbation maintains the spatial symmetry of the model which we find in figure \ref{fig:spacesymmetry}.

We find that, in general, our algorithms take systems with axis ratios (computed by comparing the radii at which, as a function of angle, the density profile reaches a value such that half the mass of the system is enclosed) of, for example, $(1.0000, 1.0005, 1.0005)$ and, after 25 perturbations, return systems with axis ratios of $(1.0000, 0.9945, 1.0167)$. In rare instances where very dispersed systems undergo rapid collapse statistical variance in the particles positions can lead to the development of triaxiality (see the deep MOND simulations where in only 5 perturbations the axis ratios change from $(1.0000, 0.9991, 0.9992)$ to $(1.0000, 0.8681, 1.3851)$). In these instances the systems still possess clear radial profiles which shall be discussed but detailed discussion the triaxiality will be avoided since at present it is not of much relevance here.

Now we need to check that the systems do not develop a preferred axis of rotation to be sure it is reasonable to speak only of radial and tangential velocity. Since the vast majority of systems retain spherical symmetry there is no issue in defining $v_r$ unambiguously leaving only the question of how to treat $v_t$. Since we are explicitly evaluating $v_{\theta}$ and $v_{\phi}$ to calculate the anisotropy we are forced to define a $\sigma_t^2$ anyway. To do so for a non spherical system is perfectly valid and will convey all the information it needs to regardless of the system's sphericity or preferred axis of rotation.

\subsection{Effect of initial anisotropy profiles}
Equation \ref{eqn:attractor} cannot be rearranged as a linear relationship between $\beta$ and $\gamma+\kappa$, yet this is the parameter space in which the data is most easily visually interpreted. Consequently, it is awkward to use equation \ref{eqn:attractor} to plot the attractor in that parameter space despite that being easiest for the reader. Accordingly, rather than derive the attractor directly from equation \ref{eqn:attractor}, we plot data from an exemplar converged solution from HJS as that is visually cleaner. This is why the attractor solution appears as a series of data points rather than a smooth, continuous function or allowed region. We first look in detail at the simplest of our initial conditions, the isotropic Plummer sphere in Newtonian gravity, expecting our data to converge around the attractor.

\begin{figure}
\includegraphics[width=84mm]{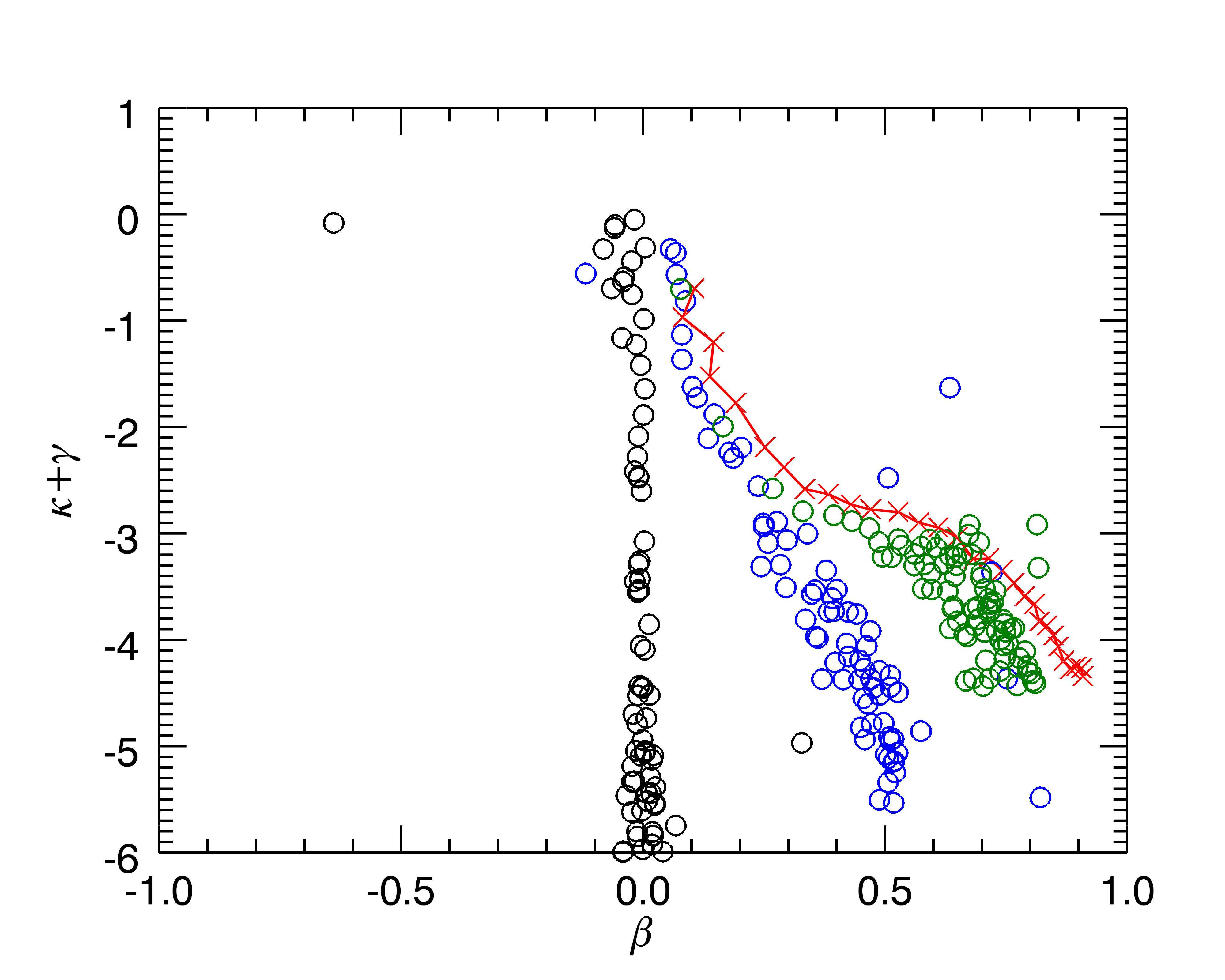}
\caption{\label{fig:plumisonewt} Evolution of the anisotropy profile of the Newtonian, isotropic initial conditions at successive (0-black, 5-blue and 30-green) `shock-flow' cycles with the attractor in red.}
\end{figure}

In figure \ref{fig:plumisonewt} the evolution of the system can be clearly seen. The initial condition is isotropic and is therefore a vertical line on $\beta=0$ with subsequent shocks apparently converging on a final state. The movement towards the end state is convergent as successive shocks disrupt the previous equilibrium less and less until, at a point around 20-30 shock cycles, the system reaches a stable state.

In interpreting plots such as figure \ref{fig:plumisonewt} a useful rule of thumb is that higher values of $\gamma+\kappa$ correspond to the inner regions of the system and more negative ones to the outer edges. This is not an exact relationship due to statistical noise and the opposing gradients of $\rho(r)$ (increasingly negative) and $\sigma_r^2(r)$ (increasingly positive), but can still be useful as an approximation due to the domination of the $\gamma$ term. This tells us that the changing anisotropy becomes apparent first in the outer regions of the system before progressively spreading inwards.

Next we consider whether, given that the attractor drives an anisotropy gradient, will an initially anisotropic model affect the evolution towards the attractor?

\begin{figure}
\includegraphics[width=84mm]{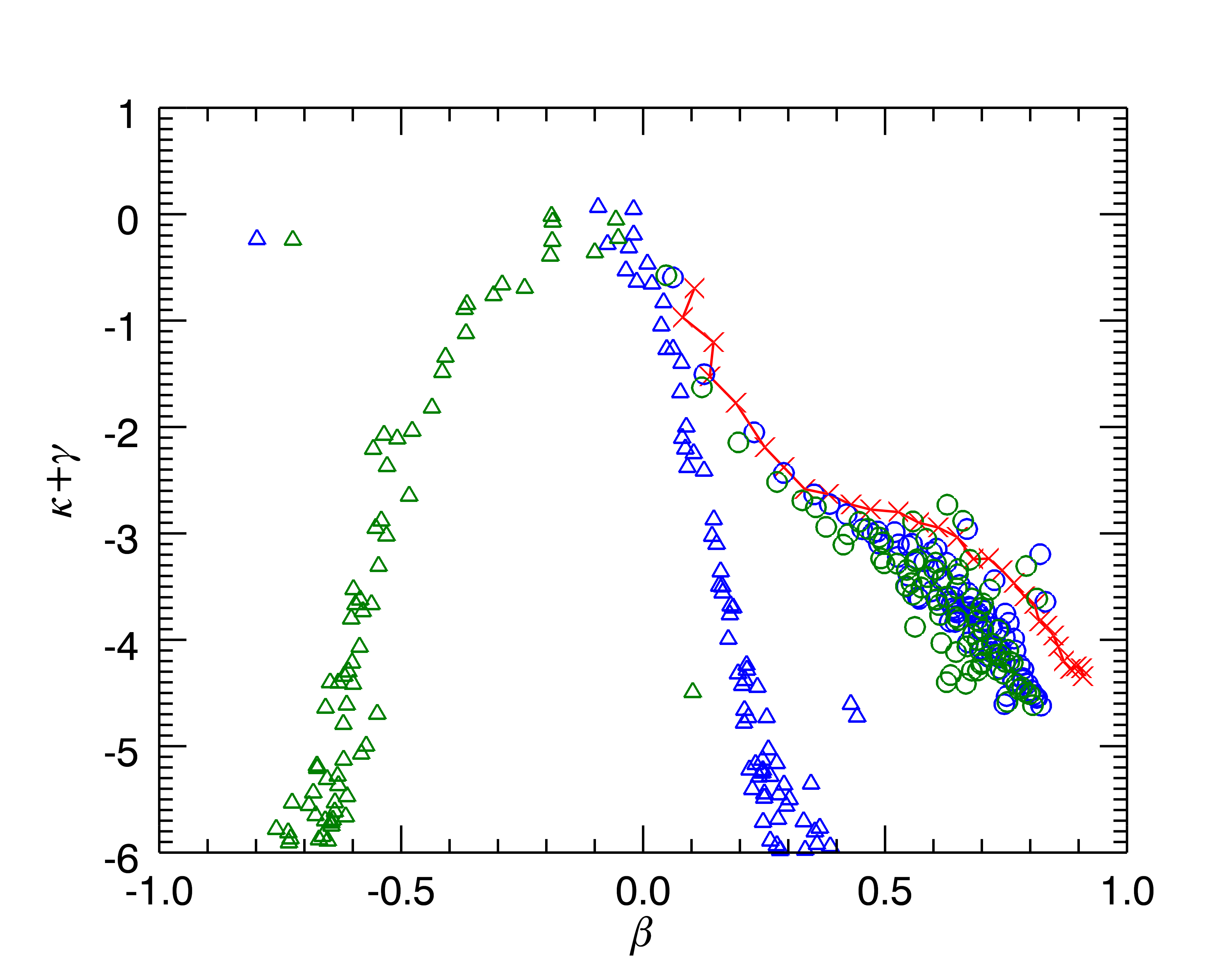}
\caption{\label{fig:plumradnewt} Initial (triangles) and final (circles) states for radial (blue) and tangential (green) initial conditions.}
\end{figure}

The anisotropy in the initial conditions is clearly visible in figure \ref{fig:plumradnewt} as a gradient in the initial data points but again the convergence is very apparent. This implies that the emergence of the attractor is robust to the initial character of the halo which is a feature that any explanation of the universal profile needs to have. Tangential models take longer to converge as the change in anisotropy needed is greater.

\begin{figure}
\includegraphics[width=84mm]{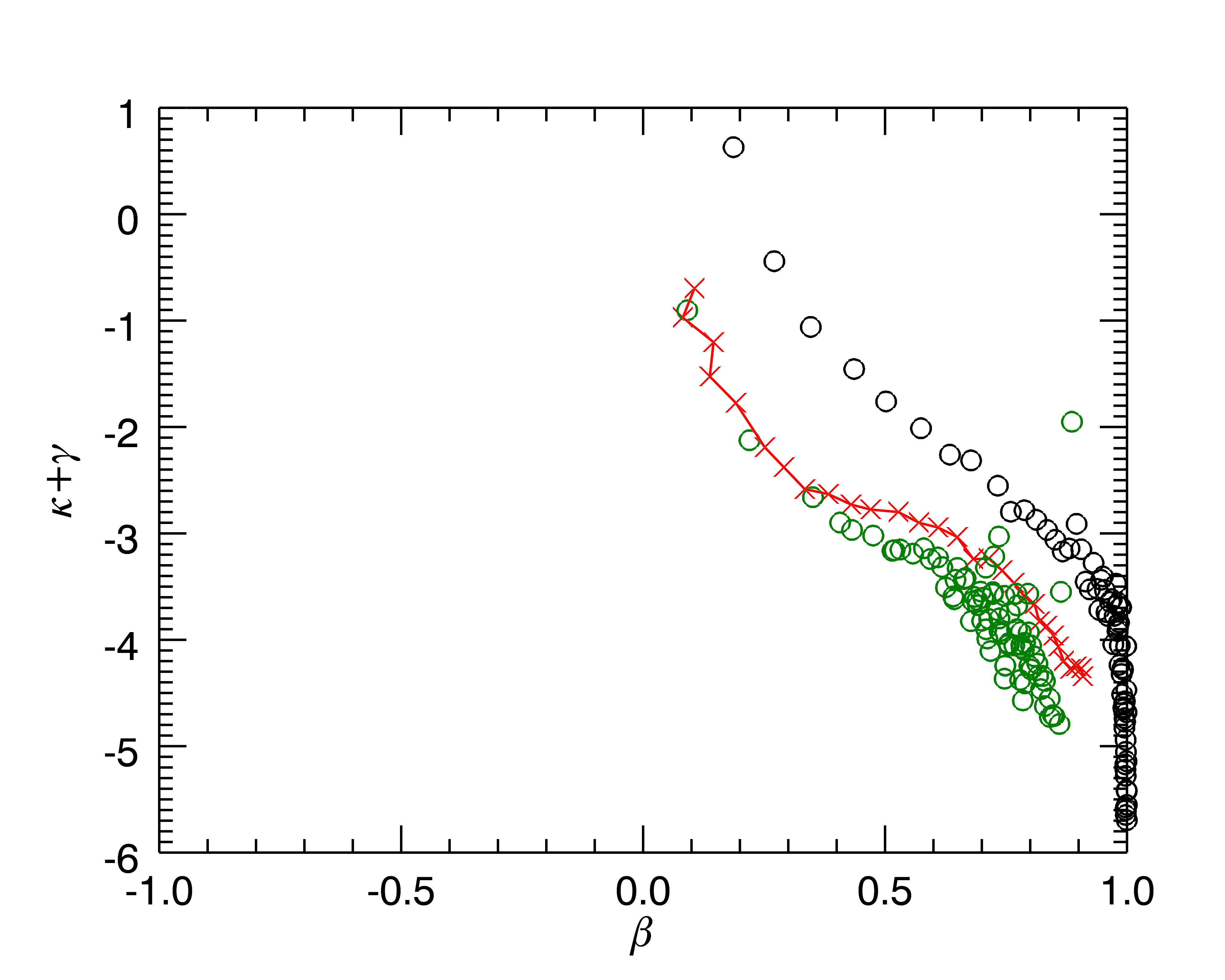}
\caption{\label{fig:s8} Initial and final states for states that lie above the attractor}
\end{figure}

Finally we try conditions that are significantly more radially anisotropic than the attractor at any given radius. These solutions also converge to the attractor just as the others do, as shown in figure \ref{fig:s8}.

\subsection{Effect of gravity theory}
Simulations were run where the initial conditions and subsequent evolution of the system were carried out under a different assumed theory of gravity during the flow phase. If the attractor is a phenomenon strongly tied, as one might reasonably expect, to gravitational laws then changing the form of those laws may have some impact on the evolution of the system.

\begin{figure}
\includegraphics[width=84mm]{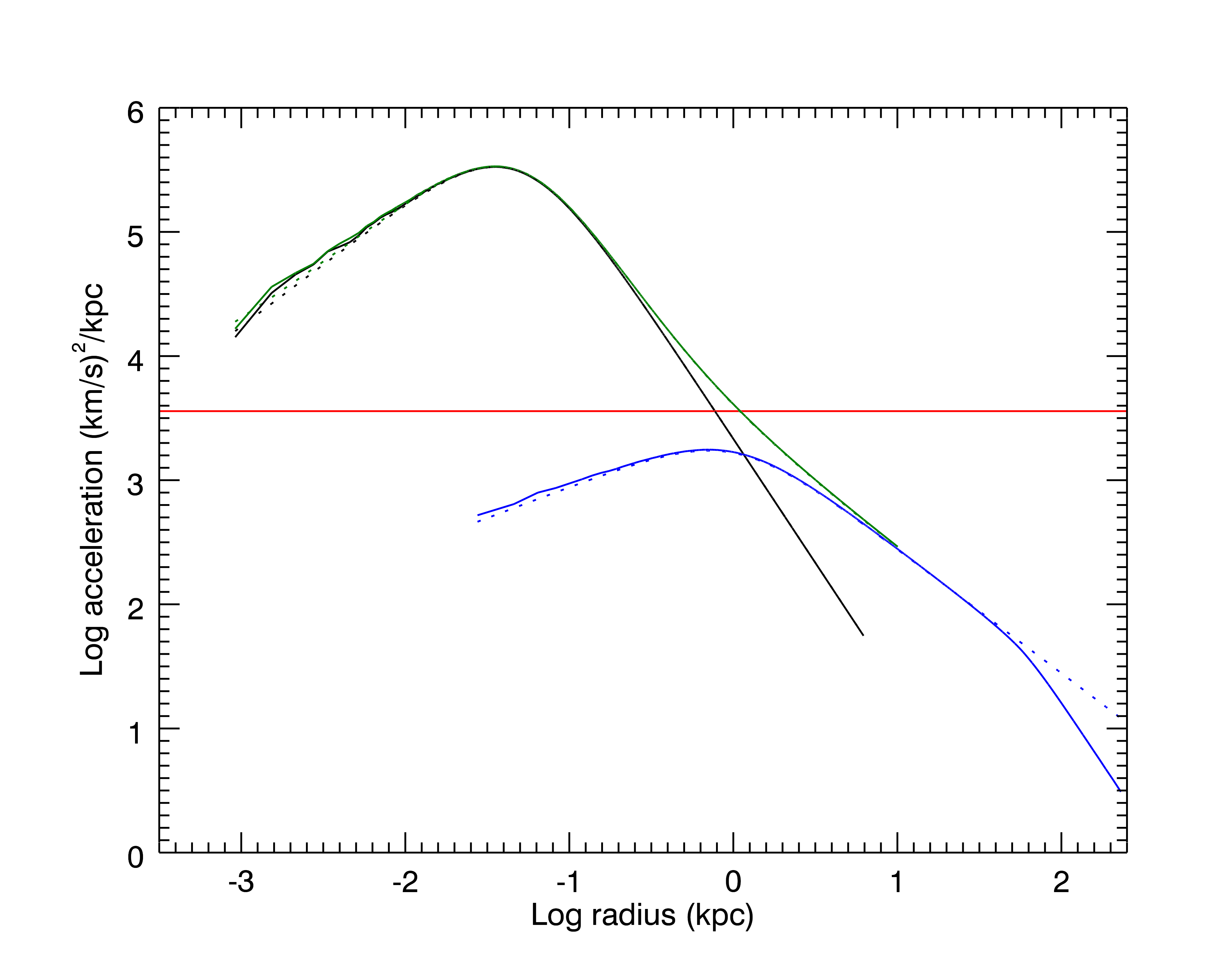}
\caption{\label{fig:mondpotential} Initial curves of acceleration (solid lines) with radius for Newtonian (black), `perturbative' MOND (green) and deep MOND (blue) models compared to their theoretical values (dotted lines) as set out by appendix \ref{sec:MOND}. Below the red line represents the region in which the MOND effects become roughly of order unity ($a < 3600 (km/s)^2/kpc$) corresponding to the deep MOND region.}
\end{figure}

We test two instances of MONDian dynamics using the same perturbations as before; a perturbative case where systems of comparable scale to those seen so far are simply translated to their stable MOND equivalent with MONDian effects only affecting the outer regions and a more extended system of the same mass where the entire system is in the MONDian regime (see figure \ref{fig:mondpotential}).

\begin{figure}
\includegraphics[width=84mm]{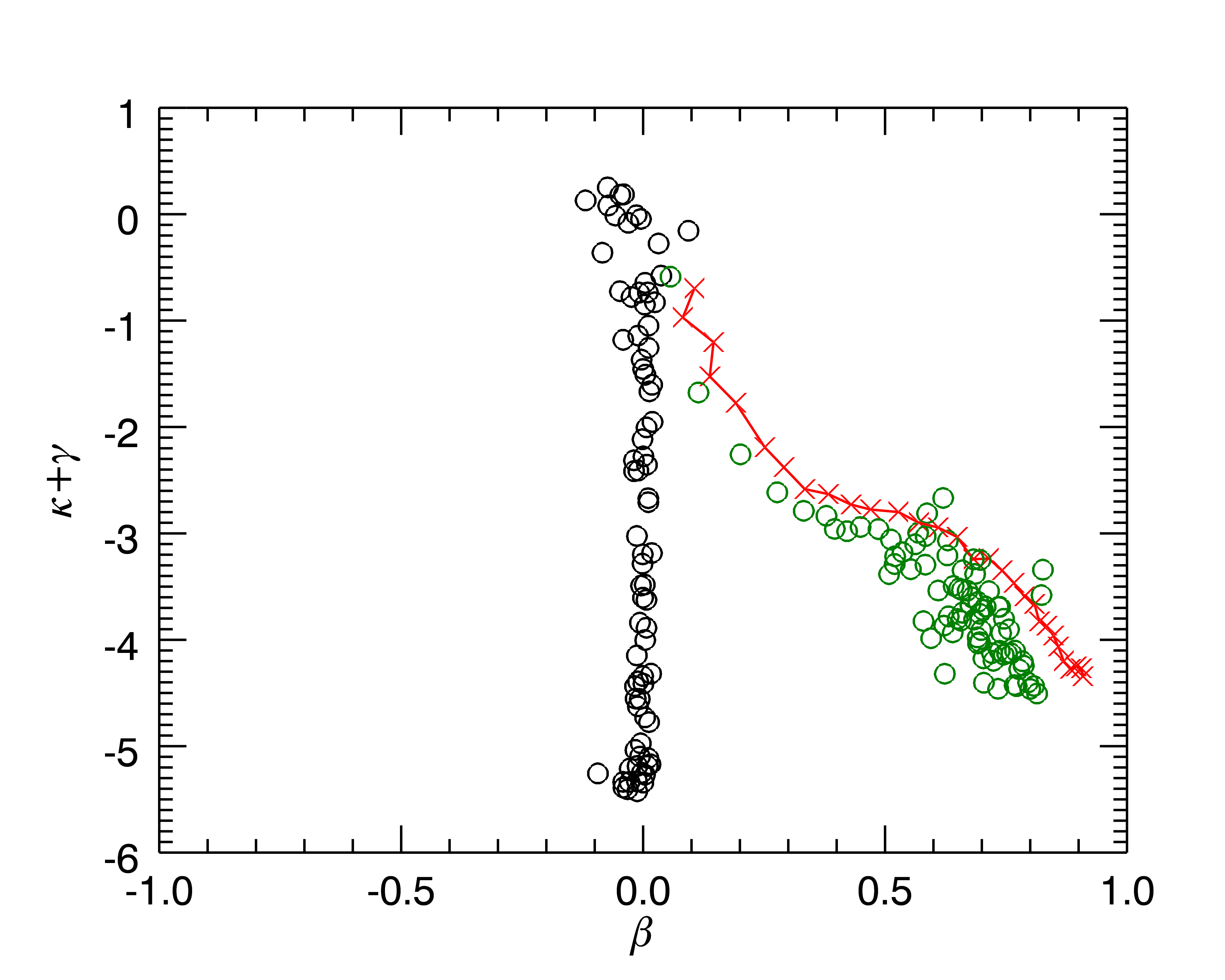}
\caption{\label{fig:plumisomond} The isotropic perturbative MONDian model.}
\end{figure}

\begin{figure}
\includegraphics[width=84mm]{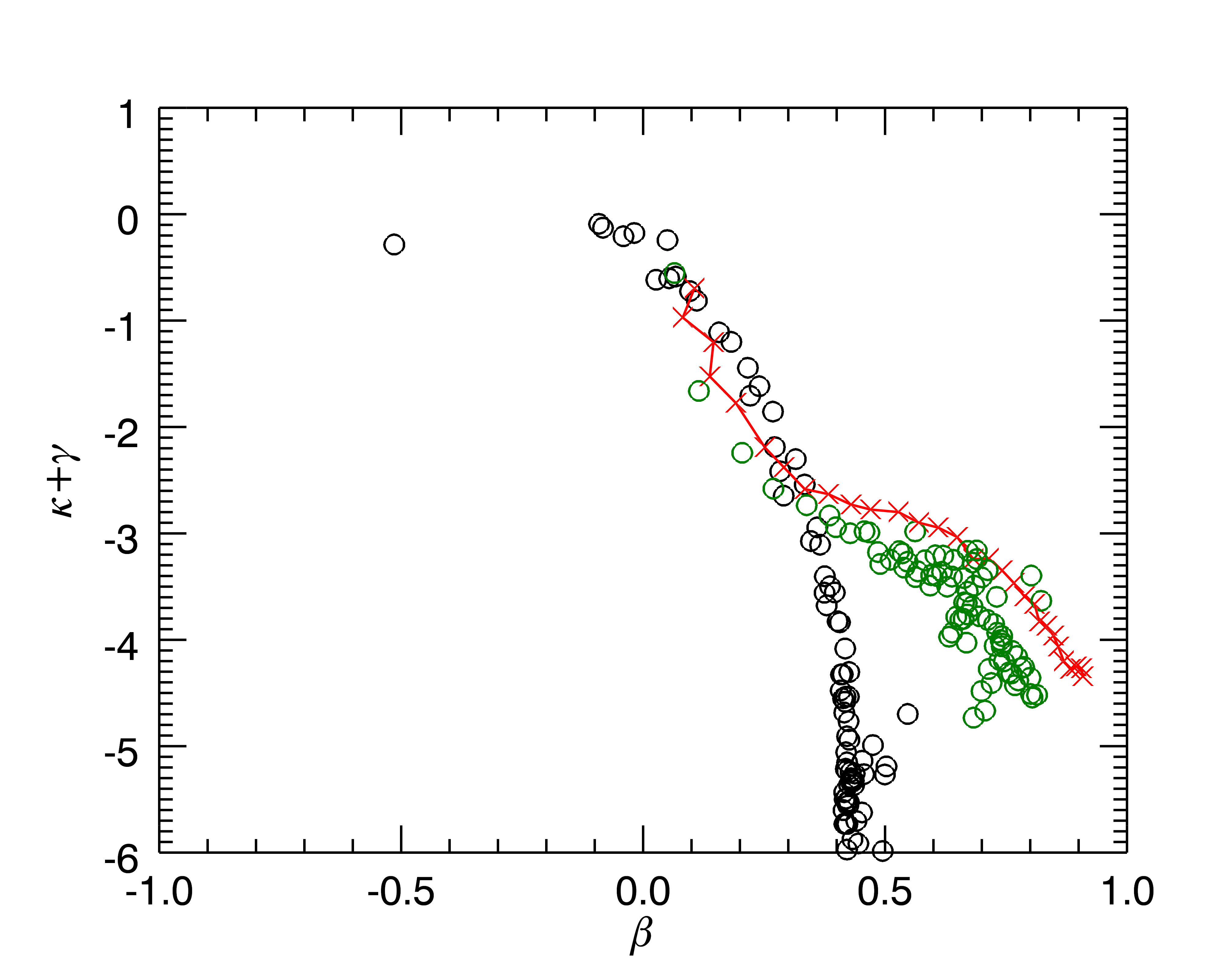}
\caption{\label{fig:plumradmond} The radial perturbative MONDian model.}
\end{figure}

The change to a MONDian context seems to have no noticeable impact on the evolution of the system as both the isotropic and radially anisotropic models, seen in figures \ref{fig:plumisomond} and \ref{fig:plumradmond} respectively, converge. The convergent solution is the same as in previous Newtonian simulations and emerges over the same timescale which is not unexpected since the MOND effect in these systems is only slight.

\begin{figure}
\includegraphics[width=84mm]{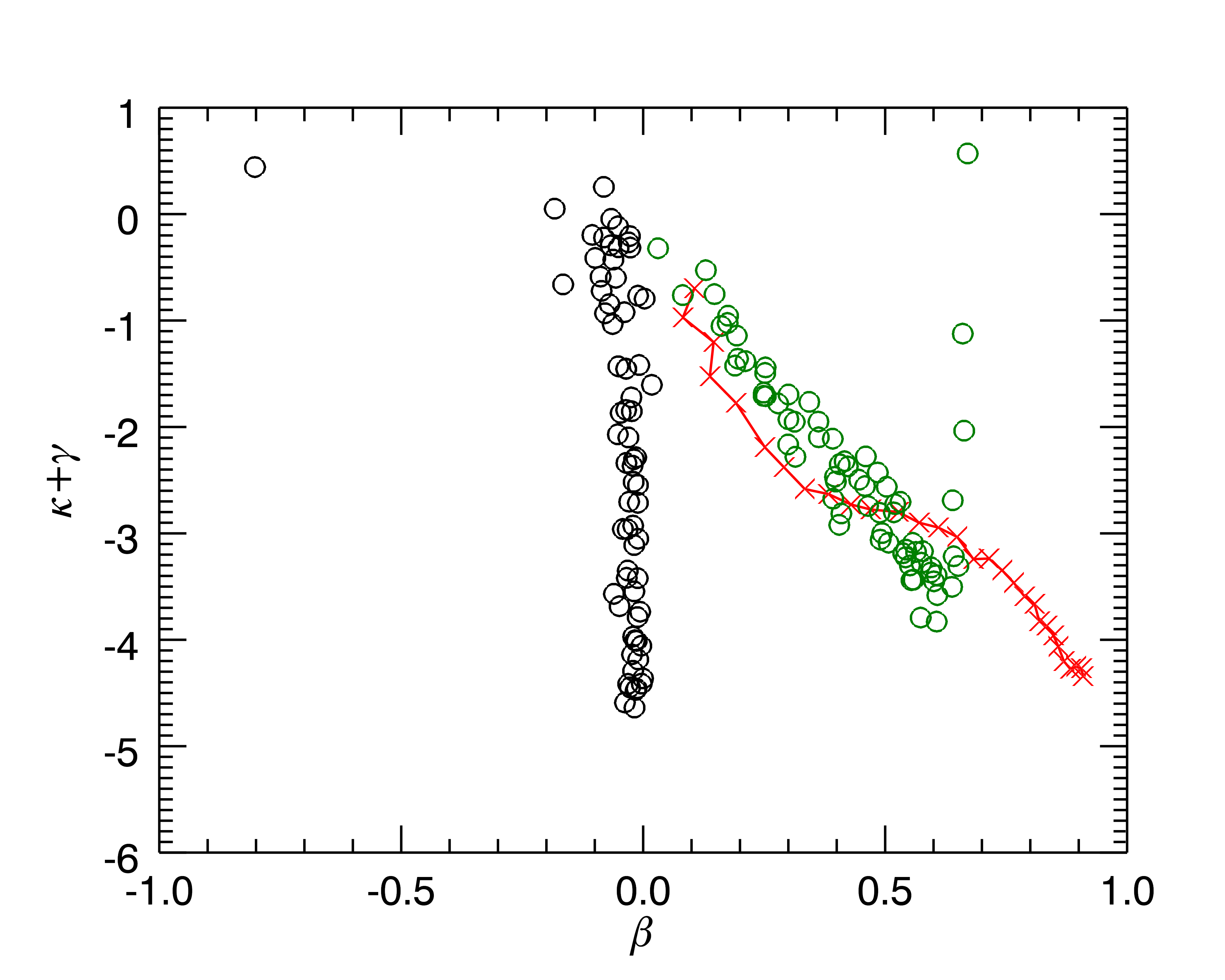}
\caption{\label{fig:plumdeep} The deep MONDian model.}
\end{figure}

When we look at the behaviour of the larger, deep MOND system in figure we still see evidence of movement towards the attractor as shown in figure \ref{fig:plumdeep}. This plot is not as clear as previous ones due to both increased statistical noise due to larger characteristic radii and the collapse of the system leading to triaxiality as mentioned in section \ref{sec:sphersym}. The collapse is interesting in its own right and is present to some extent in all systems perturbed in this manner. This will be discussed in more detail in section \ref{sec:mechanism}.

\begin{figure}
\includegraphics[width=84mm]{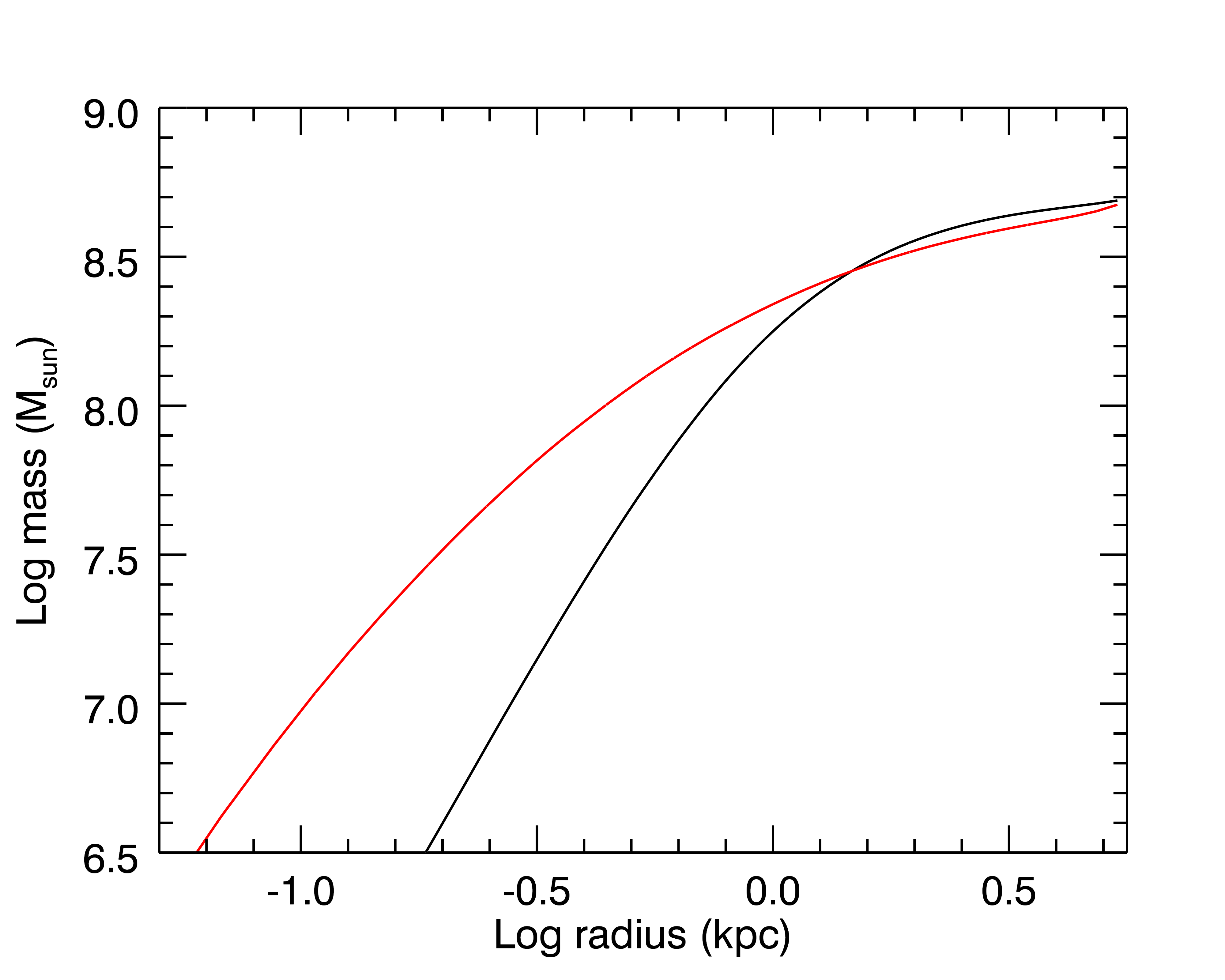}
\caption{\label{fig:collapse} The collapse of the deep MOND system after only 5 perturbations (red) compared to the initial mass profile (black).}
\end{figure}

We would not expect that the deep MOND regime would cause any significantly different behaviour from the weak MOND regime due, once more, to the tendency of our perturbation method to cause a degree of contraction in the system to which it is applied. In this case the collapse concentrates the mass and most of the mass in the system quickly leaves the deep MOND regime as, as figure \ref{fig:collapse} shows, the system becomes more concentrated.

We see that there is remarkable agreement amongst the simulations as they have all converged to a single solution. We see clear evidence for convergence to the same solution as that found by HJS. This demonstrates that the details of the gravitational interactions between particles are not especially important to the attractor. This makes sense when viewed the same way as the HJS results whereby the flow allows for the phase-mixing of the system. One would not expect small changes to gravitational coupling to have any significant impact on the system's ability to phase-mix and, indeed, this is what we find. Now we attempt to disturb the convergence by altering the algorithm.

\section{Altering the algorithm}

\subsection{Implicit coordinate system}
The perturbation algorithm that we use has dealt with velocity vectors in a purely cartesian way however we want reassurance that the attractor is independent of the coordinate system. The Newtonian, isotropic system was re-run but this time the random scaling of velocity vectors $v_{[x,y,z]}$ was carried out on $v_{[r,\theta,\phi]}$ instead. When the velocities are converted back to into Cartesian format the scaling will no longer be uniform or in the same limits as before thanks to the interconnected and non-linear nature of the transforms used e.g.:

\begin{equation}
\label{eqn:coord}
\dot{r}=\frac{x\dot{x}+y\dot{y}+z\dot{z}}{\sqrt{x^2+y^2+z^2}}\rightarrow \dot{x}=r\left(\cos(\theta)\cos(\phi)\dot{\theta}-\sin(\theta)\sin(\phi)\dot{\phi}\right)
\end{equation}

We find that the move from Cartesian to spherical coordinates has no effect on the convergent solution. This is in accordance with the findings of HJS where they tested different limits on the magnitude of the random factor used to scale the velocities.

\subsection{Random scale factors and flow time}
\label{sec:flowtime}
\begin{figure}
\includegraphics[width=84mm]{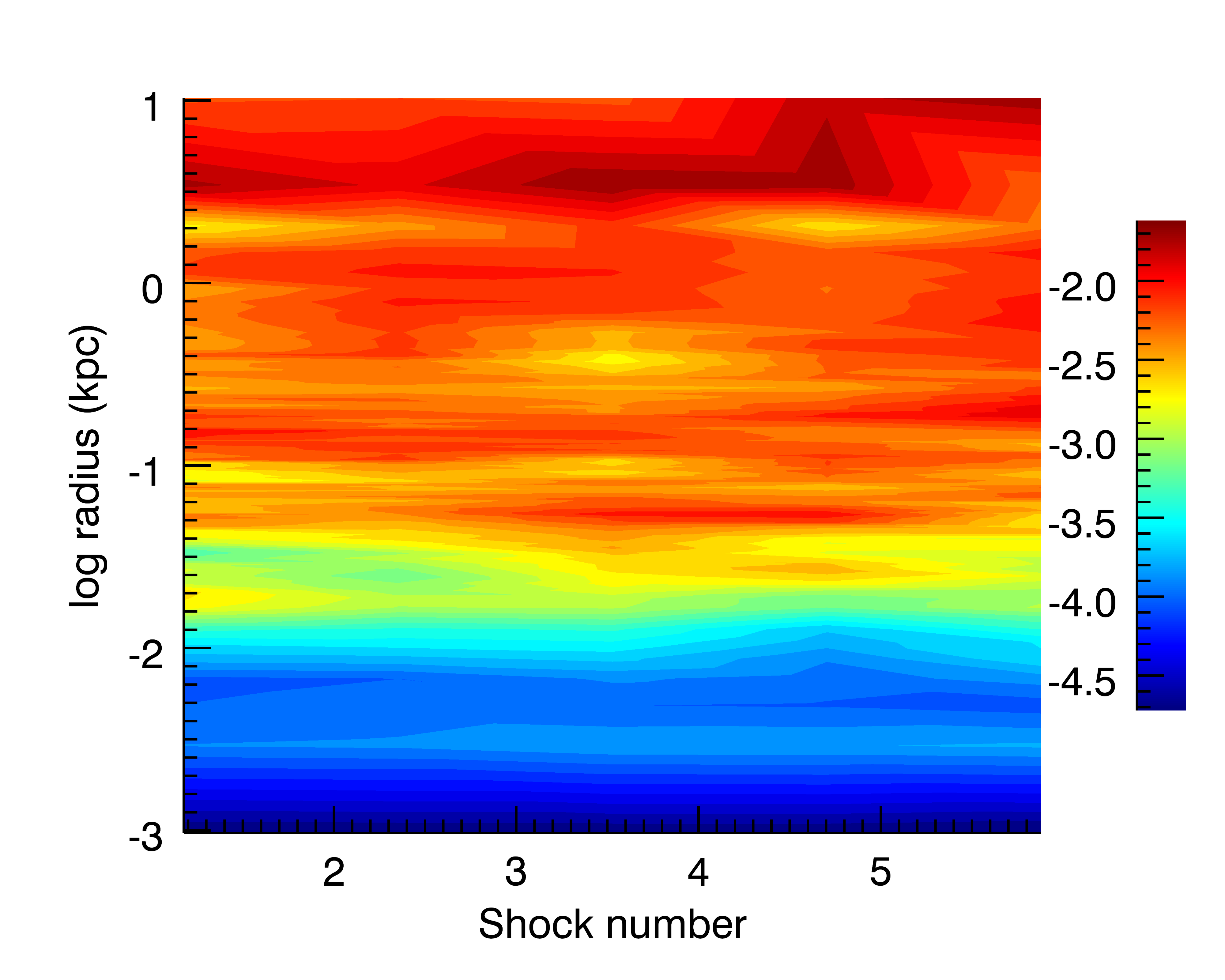}
\caption{\label{fig:flowcontrol} The changes in anisotropy in a stable, unperturbed system. Note that since the changes in anisotropy are expressed clearest as logarithms the change being evaluated is $\log{|\beta_f-\beta_i|}$. The `shocks' listed on the x-axis are only marked for easy time comparison with plots where shocks \emph{were} actually applied. Here the shocks simply mark intervals of 1 $T_{dyn}$. \emph{Note the difference in scale from figure \ref{fig:flowtest}}}
\end{figure}

In order to talk quantitatively about the speed of convergence and to demonstrate that the systems have actually reached a convergent solution we need some way of defining convergence. Since the evolution of the system is most clearly seen in the changing anisotropy profile we choose to quantify the convergence by comparing the anisotropy profiles before and after a kick-flow cycle and comparing that with the statistical drift that one would expect to see in the anisotropy profile of a system that is in equilibrium. This is visualised in terms of a plot such as figure \ref{fig:flowcontrol} where, at every radius, the change in anisotropy between successive steps is plotted and then smoothed into contours to show the convergence. Figure \ref{fig:flowcontrol} is a control plot which shows the results of such a plot on the isotropic Newtonian IC's when no shocks are applied. Note that since the changes in anisotropy are expressed clearest as logarithms the change being evaluated is $\log{|\beta_f-\beta_i|}$.

Figure \ref{fig:flowcontrol} shows us that the equilibriated central regions are very stable with fluctuations on the scale of $\pm10^{-4}$ increasing to $\pm10^{-2}$ towards the edge of the system. Thus we can define convergence in terms of a change in beta that is less than or equal to the change at the same radius in the unperturbed system, figure \ref{fig:flowcontrol}.

In practice, we would not expect such precise stability. The attractor does present a favoured configuration in one parameter space but, as we shall demonstrate late, Jeans' equation means that a variety of density profiles and dispersion will fit it. Since the analysis performed to produce these plots uses bins of fixed radius the subtle changes in the density profile are a source of noise that is not present in the control plot. In practice, one uses these plots to identify when the changes in anisotropy are roughly constant at a given radius (since the rate of convergence is not constant normally) and when the fluctuations in anisotropy are reasonably close to those in the control plot.

\begin{figure}
\includegraphics[width=84mm]{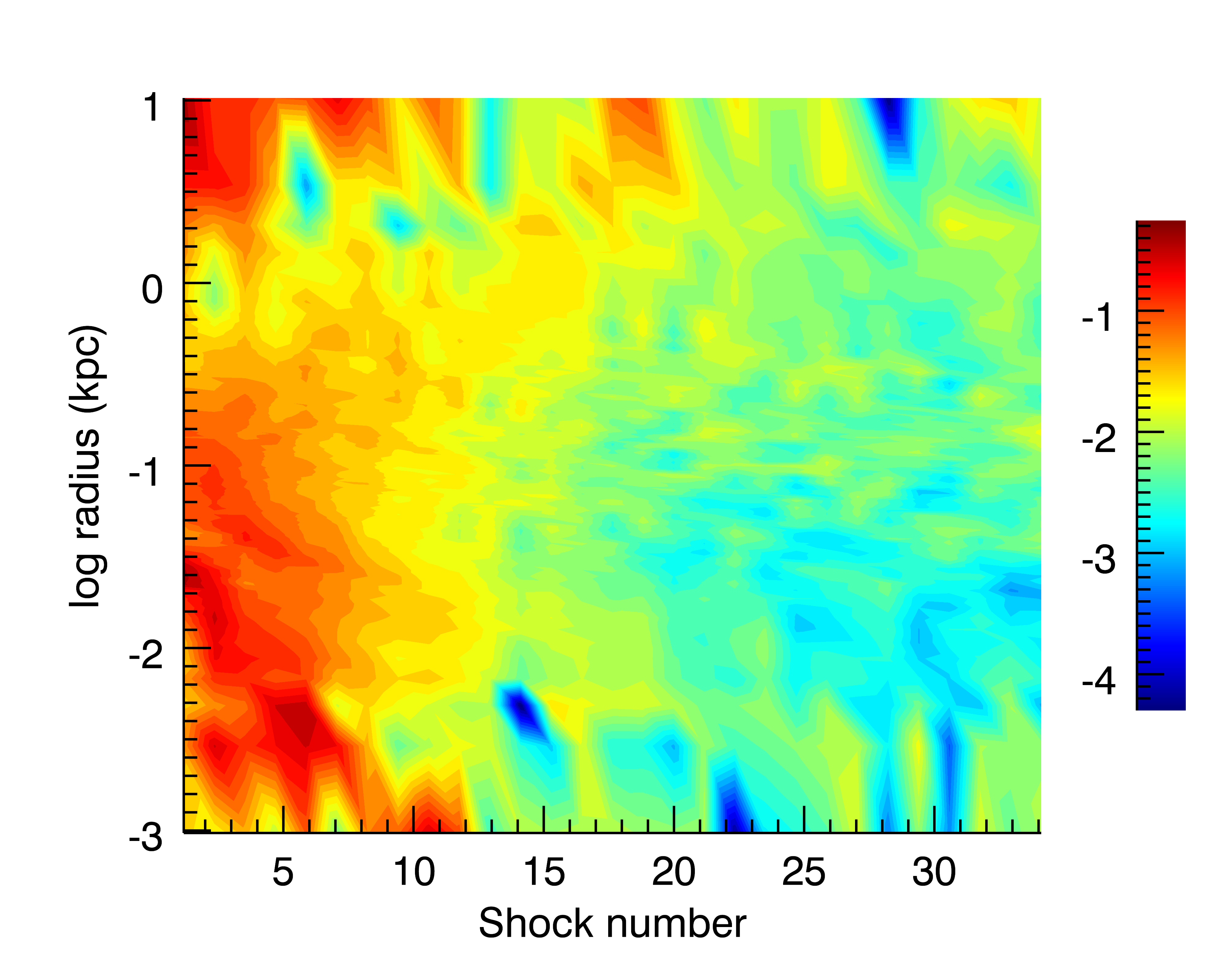}
\caption{\label{fig:flowtest} The changes in anisotropy over 30 kicks of the Newtonian, isotropic system. \emph{Note the difference in scale from figure \ref{fig:flowcontrol}}}
\end{figure}

Noting that the colour scales of the two plots are different, we see that the system is actually very stably converged. Fluctuations in the outer edges of the system are very close to the control plot while the central regions are a little more unstable inside of $0.01\ kpc$, with moments when the system is as stable as the control and moments when it is not. This is taken as reasonable indication that the system has evolved as far as it is going to. Similar plots for our other systems demonstrate that they are also fully converged.

As noted previously there is a lot of freedom in designing perturbation algorithms with particular freedom allowed in the numerical ranges of certain critical parameters such as the range of the scaling factor and the length of time allowed for the flow phase. In order to examine the effect of parameter choices on the progression of our simulations we ran four simulations using different combinations of scaling factors and flow times and then stopped them after only a few perturbations. We can then see which simulations have evolved further in that time and which are slower. We stop the simulations before they reach the attractor because at that point all the simulations should be lying on top of each other anyway and any differences in how they progressed to that point would be lost.

\begin{figure}
\includegraphics[width=84mm]{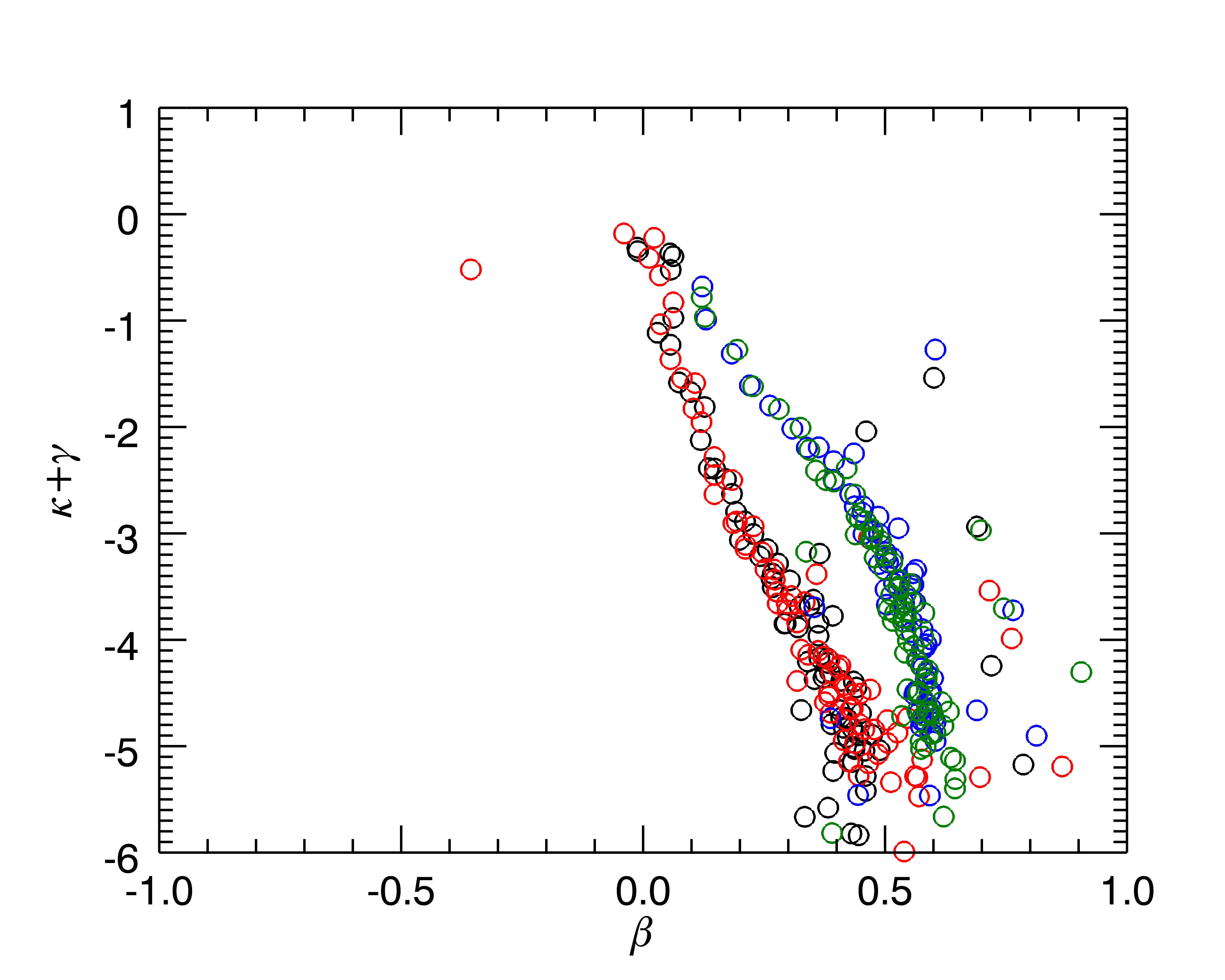}
\caption{\label{fig:ratecomp} Attractor plots for four different combinations of algorithm parameters: Using scaling factors of $0.5<f<1.5$ with 1 (black) and 3 (red) dynamical times for the flow and $0.0<f<2.0$ with 1 (blue) and 3 (green) dynamical times. Each simulation has only been perturbed 4 times by this point. This number is arbitrary and was chosen as it presented the clearest plot.}
\end{figure}

Looking at figure \ref{fig:ratecomp} we can see why it is important to thoroughly examine the choice of algorithm as some parameters have a significant impact while others do not, even if one might expect otherwise. For example, we show here that allowing the system to equilibriate for longer periods of time actually has almost no effect on the speed of convergence. This is slightly counter to expectation but amply demonstrates both the stability of all the intermediate solutions and the fact that the attractor does require the system to be driven to it to some degree but that that the final point of convergence is special compared to all the intermediate steps. In our scheme we see that the most important influencing factor is the size of the range of scaling factors applied with larger kicks promoting faster convergence.

\section[]{Mechanism for convergence}
\label{sec:mechanism}

\subsection{One shock cycle in detail}
\label{sec:detail}
\begin{figure}
\includegraphics[width=84mm]{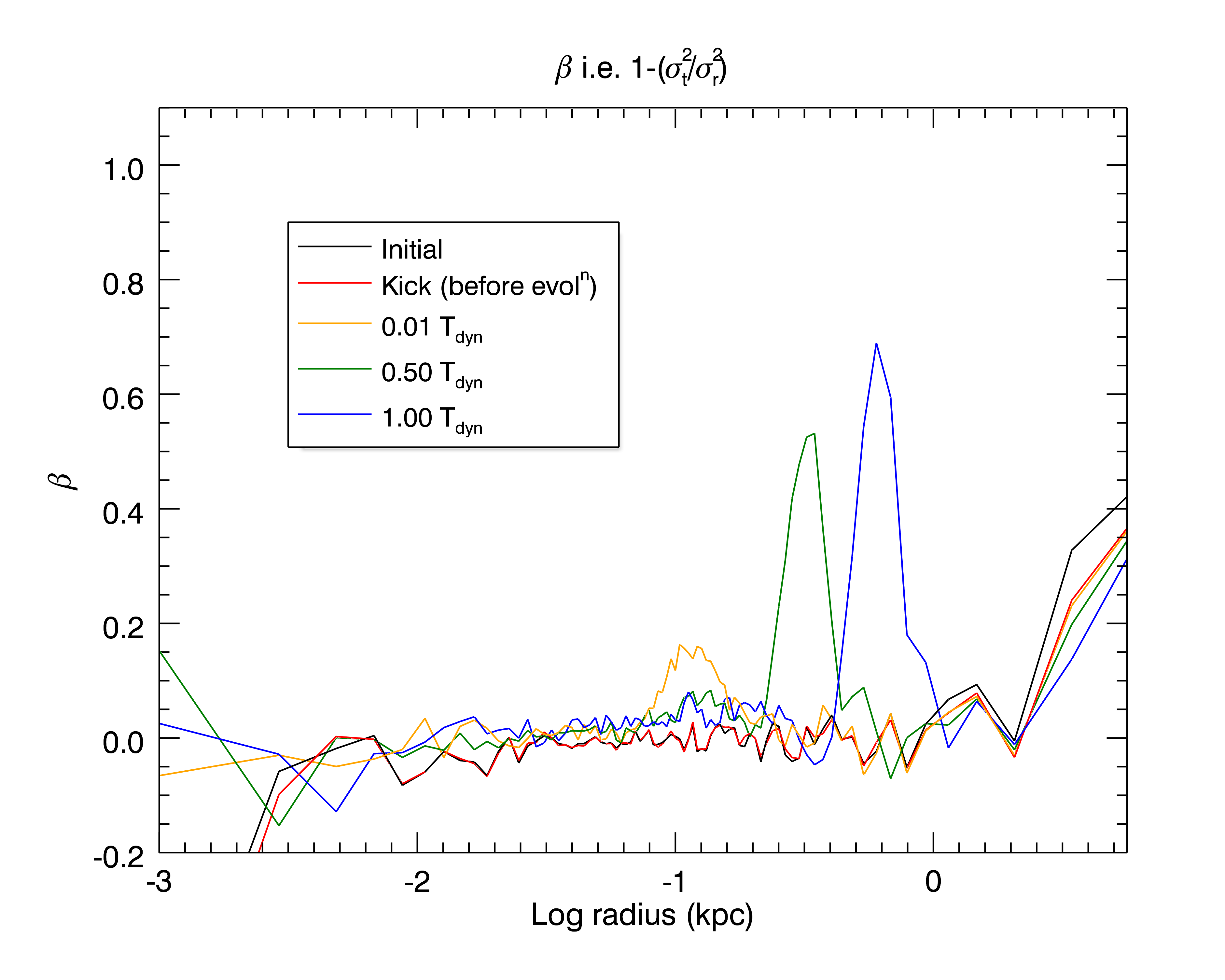}
\caption{\label{fig:betasingle} The anisotropy profile of a standard isotropic system during the initial conditions, after the kick has been applied but before the system has evolved and then subsequent profiles as the system relaxes.}
\end{figure}

In order to understand the development of the convergent behaviour we investigate the changing state of the system during a single flow phase. We see several notable changes that progress over the course of the flow. The most interesting effect is the propagation of the anisotropy which seems to start in the inner regions and then subsequently travel outwards as a wave of radial anisotropy which can be seen in figure \ref{fig:betasingle}. This is because radial anisotropy denotes a population of particles with unusually high radial velocities which means that those particles will now be traveling towards the outer edges of the system.

Note that the wave is still very clear even after 1 $T_{dyn}$. However, since figure \ref{fig:betasingle} is plotted as a function of radius, there is actually very little mass between the wave and the edge of the system. This means that, if the system is left to evolve further, the profile as a function of mass will not change significantly. This is supported by figure \ref{fig:ratecomp} where allowing the system to flow for a further 2 $T_{dyn}$ has no discernable impact on the system's evolution towards the attractor. Hence we can say that these systems are as close to equilibrium as makes no difference and can, for our purposes, be treated as fully evolved and suitable for another perturbation.

The perturbation scheme itself does not instantaneously perturb the anisotropy profile to any significant degree but we see that within a very small amount of time we have a pronounced peak over a range of radii which then proceeds to migrate to the outer regions of the system. The peak originates from the highest density area of the system which follows from the statistical nature of the cause of the effect.

What this shows us is that the perturbation sets up a population of strongly radial particles but in such a way that they are not simply aligned specifically to be now traveling radially. If the latter were the case then the anisotropic signal would show up in the profile the instant after the perturbation algorithm completes whereas we see that some time, here at least 0.01 $T_{dyn}$, must pass before the anisotropy presents itself. This is supported by the aforementioned symmetry of the perturbation algorithm in its treatment of velocity vectors and the fact that the algorithm does not know what the radial velocity vector of a given particle even as it is, as we showed in the previous section, coordinate-system independent.

The cause for this radial preference is due to the statistics involved in putting a particle on a radial orbit. Consider a particle on a circular orbit around a centre of mass. If we perturb that particle using our scheme then, unless that particle is lucky enough to have it's velocity components all scaled by unity, it will either gain or lose energy and end up on a more radial orbit either now falling towards the centre of it's orbit or being ejected outwards. If we want to return our test particle onto a more circularised orbit we must now be extremely fortunate and apply just the right scaling to each velocity component so that not only does it end up with the correct total energy but it is now moving tangentially to the centre of it's new orbit. This is far less likely than it just being put on another elliptical orbit and thus contributing to the radial anisotropy.

\begin{figure}
\includegraphics[width=84mm]{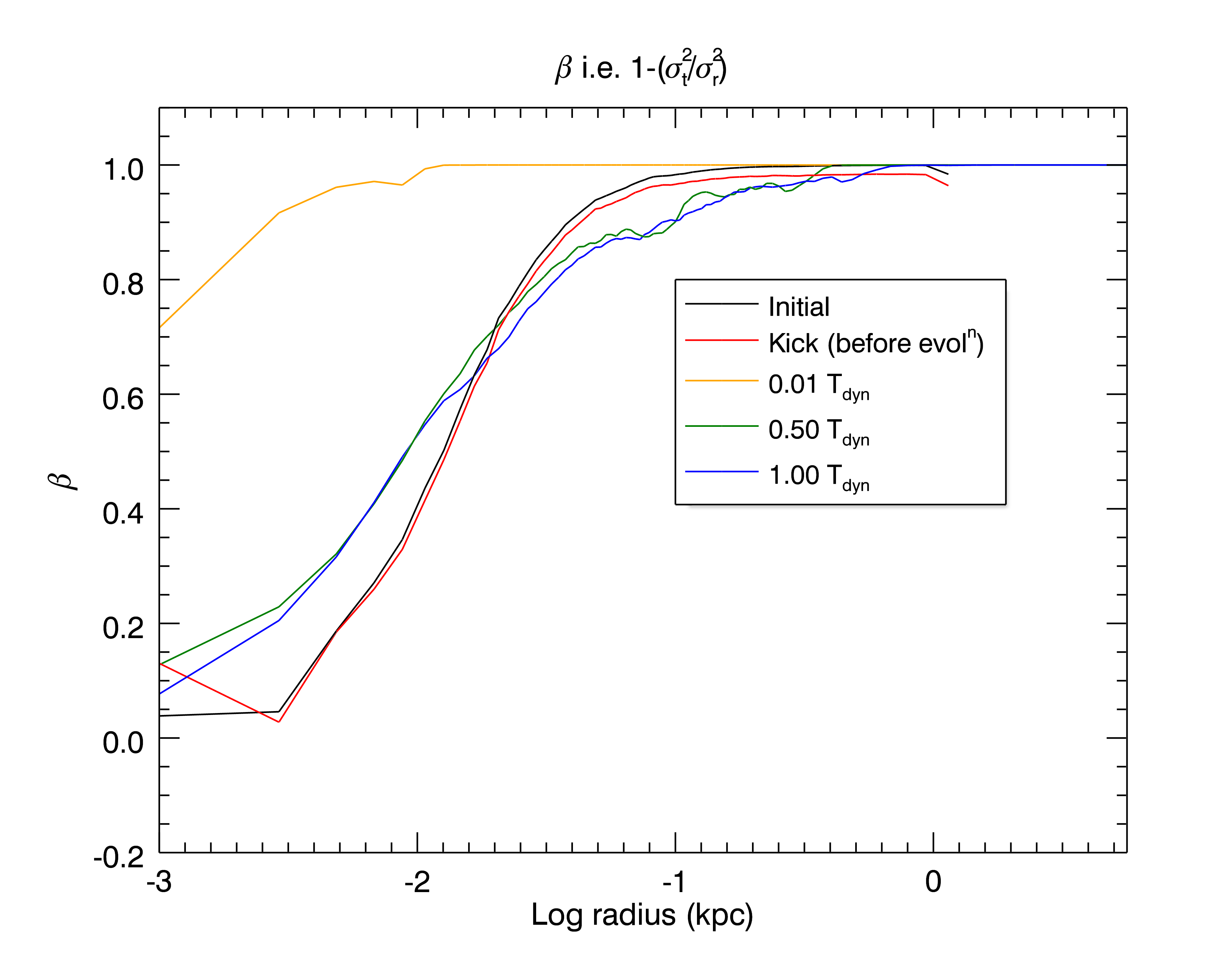}
\caption{\label{fig:s8single} The anisotropy profile of the very radial IC during the initial conditions, after the kick has been applied but before the system has evolved and then subsequent profiles as the system relaxes.}
\end{figure}

We do see different behaviour from systems that are initially more radially anisotropic than the attractor. In figure \ref{fig:s8single} we still see that the perturbation itself does not change anything significantly but we are then missing the `wave' of anisotropy visible in figure \ref{fig:betasingle}. This is because the mechanism here has more in common with standard radial orbit instability (ROI) and thus the most significant effect seen in the profile is the inward collapse of the very radial outer edges. The outward `wave' of high energy particles produced by the perturbation is still there but the lack of an isotropic background prevents it from standing out. It's presence can be inferred from the increased size of the system. Since studies of ROI show that radial populations are highly susceptible to this kind of collapse \citep{Barnes2009}, what we observe in figure \ref{fig:s8single}, is not unexpected.

The radially anisotropic outer parts of the system quickly collapse radially inwards (the yellow line from figure \ref{fig:s8single}) as the kick has robbed them of energy. As they move inwards they dramatically increase the radial anisotropy of the region they pass through until they settle into more circular orbits at smaller radii (blue and green lines). The more isotropic regions in our system behave the same as the system from figure \ref{fig:betasingle} and thus we have a general increase radial anisotropy anyway. Overall, we find the nucleus becomes more radial at the expense of the envelope thanks to the migration of the outer regions. We still see a very radial population at the extreme edge of the system as the highly energetic, almost ejected `wave' population, seen so clearly in figure \ref{fig:betasingle}, is still present, just overshadowed by the collapse of the existing radial population.

\subsection{Overall effect of the perturbation}
\begin{figure}
\includegraphics[width=84mm]{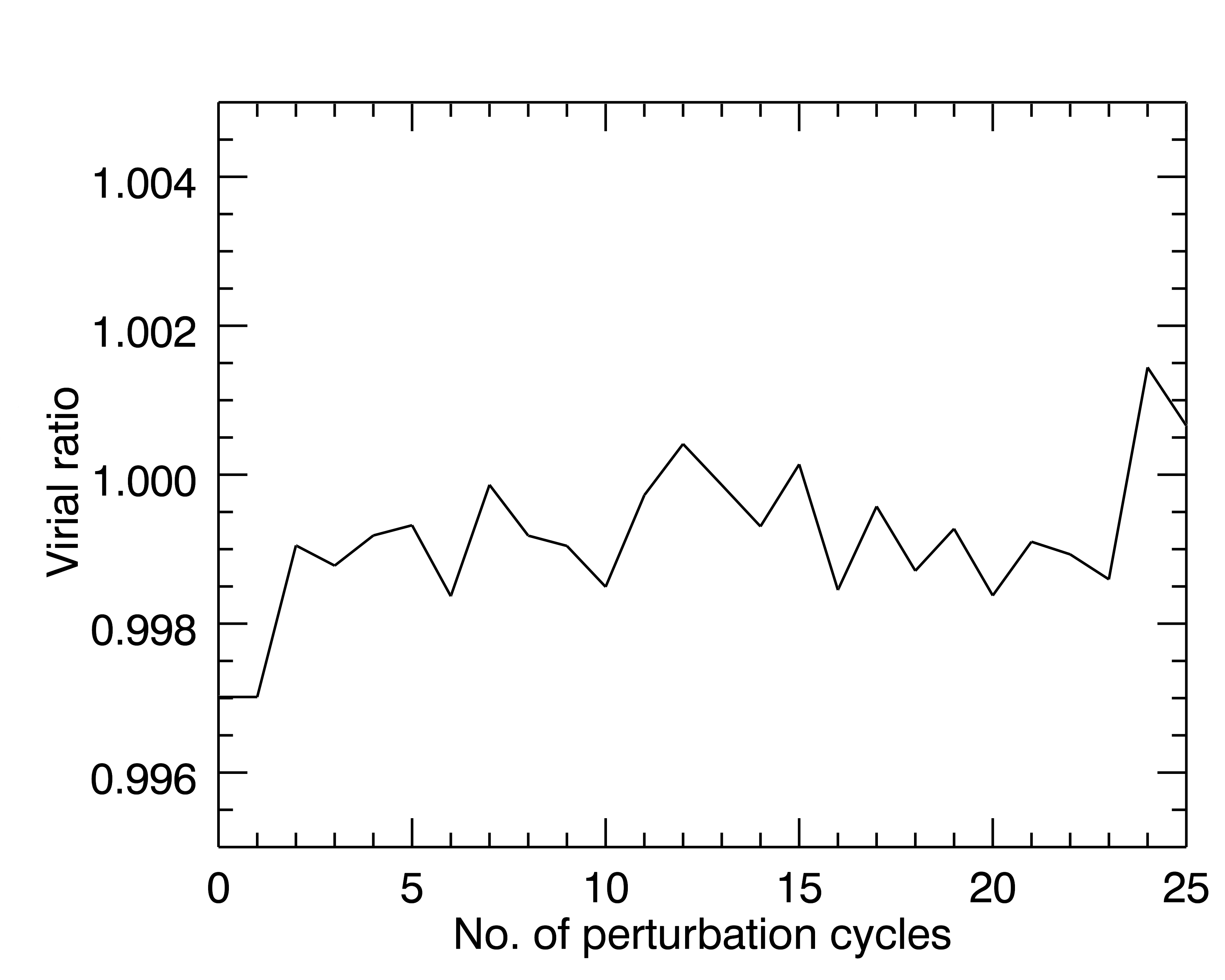}
\caption{\label{fig:virial} Virial equilibrium at the end of every flow phase for the isotropic Newtonian simulations.}
\end{figure}

While the presence of an attractor in the parameter space seems quite clear it is not obvious what mechanism is at work to evolve the system towards that attractor. We can identify some effects, such as the radial population discussed in the previous paragraph, but it does not appear that we are driving the system significantly out of equilibrium with our perturbation scheme. To within reasonable limits the system never leaves virial equilibrium. We know, from the nature of our conservation laws, that the kinetic energy is rigidly conserved and the potential energy is never affected so the start of the flow is always virialised if the preceding flow phase ended virialised. Plotting the virial ratio at the end of every flow phase in figure \ref{fig:virial} for the isotropic Newtonian case shows that the system is always in equilibrium.

In addition to this, there is no obvious analytical path to producing this kind of anisotropy. If our perturbation scheme treats each independent velocity component equally then an initially isotropic distribution should be transformed to another equivalent isotropic distribution. If we begin from the definition of $\dot{r}$ from equation \ref{eqn:coord} but re-written for clarity here as:

\begin{equation}
\label{eqn:radial}
v_r=\frac{v_xx+v_yy+v_zz}{r}
\end{equation}

What we are explicitly conserving is the energy in the system i.e. the square of the velocities:

\begin{equation}
\label{eqn:radial2}
\bar{v_r^2}=\left(\frac{v_xx+v_yy+v_zz}{r}\right)^2=\bar{v_x^2}\frac{x^2}{r^2}+\bar{v_y^2}\frac{y^2}{r^2}+\bar{v_z^2}\frac{z^2}{r^2}+\left[2\bar{v_xv_y}\frac{\bar{xy}}{r^2}+\dots\right]
\end{equation}

Since we are in an orthogonal system the cross terms are all $0$ which leaves us with:

\begin{equation}
\label{eqn:radial3}
\bar{v_r^2}=\bar{v_x^2}\frac{x^2}{r^2}+\bar{v_y^2}\frac{y^2}{r^2}+\bar{v_z^2}\frac{z^2}{r^2}
\end{equation}

So, in a large enough sample size, the random scale factors that we apply will sum to unity and we will not alter $\bar{v^2}_r$. The only distribution that we do alter is the distribution of the particles in energy momentum space where the perturbation acts to spread out the distribution with successive shocks. This can be imagined as, in each bin, the particles with the highest and lowest velocities might get scaled to even high and even lower velocities respectively which has the effect of broadening the distribution in that bin.

\begin{figure}
\includegraphics[width=84mm]{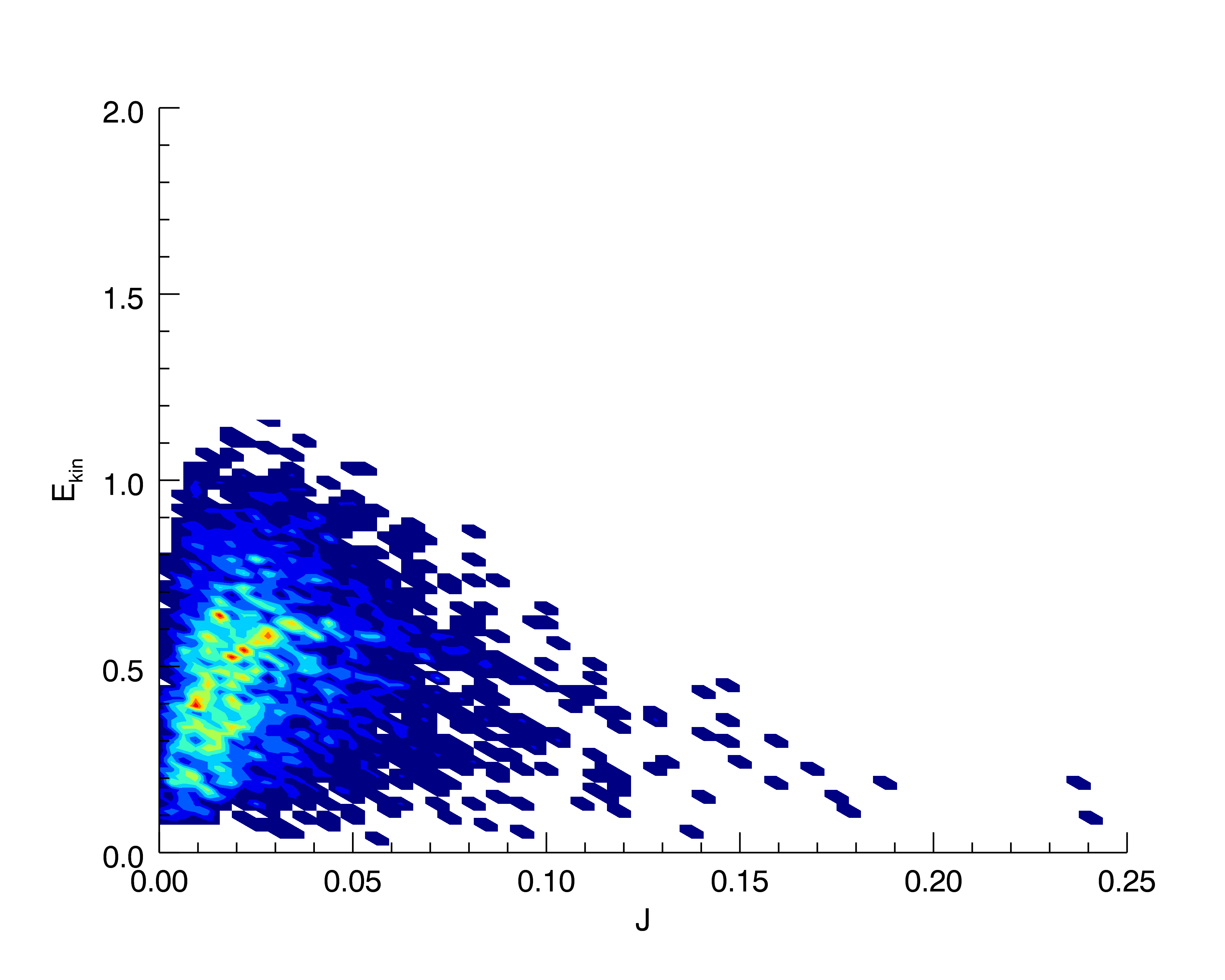}
\includegraphics[width=84mm]{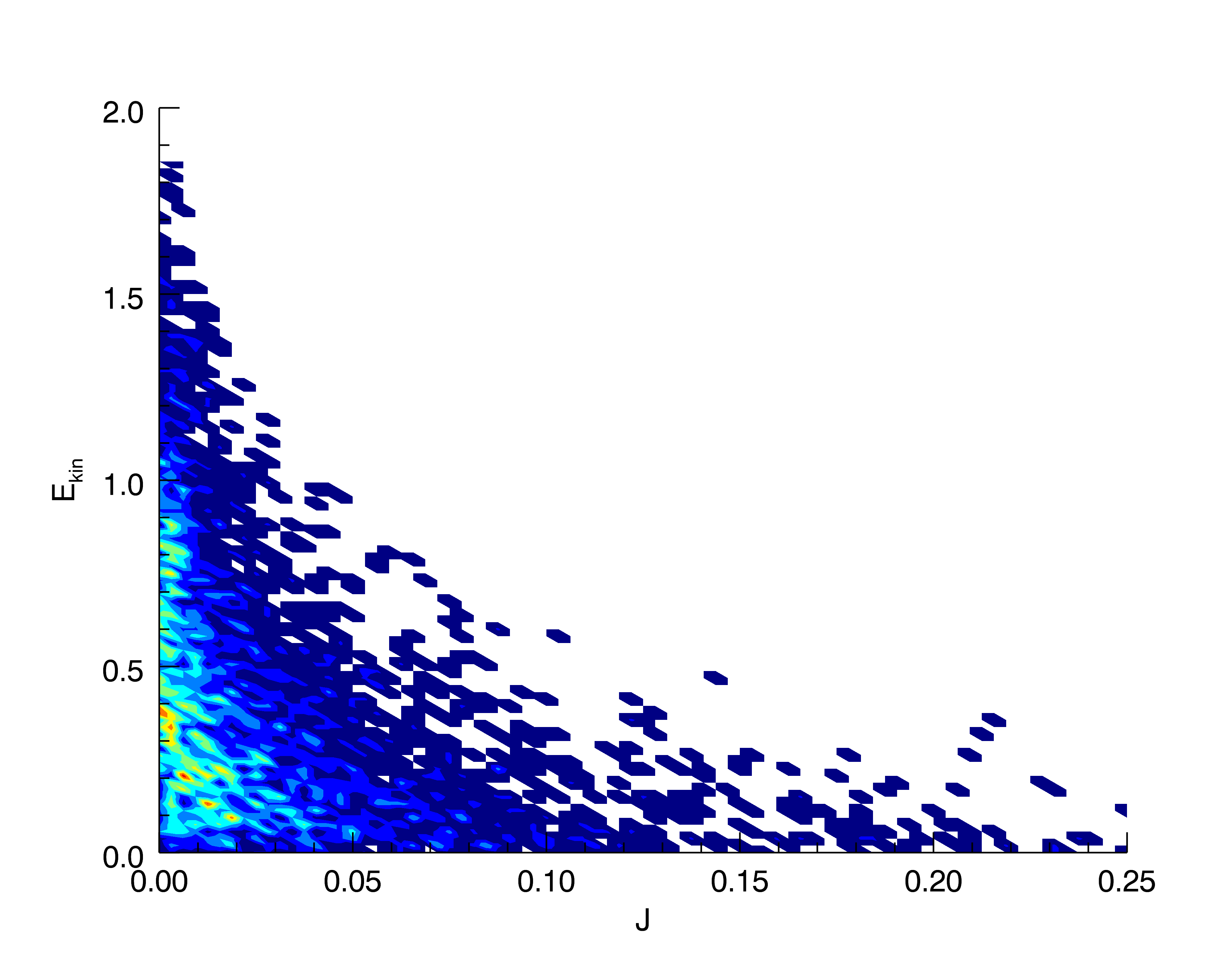}
\caption{\label{fig:EJspace} Evolution of $E_{kin}$-$J$ number density contours for the isotropic Newtonian. The hexagonal effect in low density areas is an artifact of the binning process used for the contours.}
\end{figure}

Figure \ref{fig:EJspace} shows how the kinetic energy distribution spreads out while the distribution in J becomes slightly more compacted towards $0$. The latter effect is because of the increased radial anisotropy; for a given energy in the final state, more of that energy is in the radial modes leading to a lower characteristic angular momentum for that energy. Note the clear boundary marking an excluded zone in the upper-right quadrant. This arises due to two limiting effects, both aspects of the restriction on particles leaving the system. Energy is restricted simply in relation to the local binding potential which in turn is related to radius which can be plausibly proxied by angular momentum. Angular momentum is restricted because, as touched on before, angular momentum for a given energy is limited by the momentum of a circularised particle with that energy. These two effects create a decaying cut-off defining an allowed region of values.

Another behaviour that seems characteristic of the perturbation is the splitting of the system into two populations of particles: a `nucleus' of particles who move towards the inner regions of the system and a small population of `envelope' particles which move further out. This is most obvious in the larger, more extended systems as the collapse of the larger, `nucleus' population is very noticeable whereas in most simulations it is the `envelope' population that makes its presence most apparent as the increase in physical size increases the dynamical timescale and thus the time the simulation must run for.

\begin{figure}
\includegraphics[width=84mm]{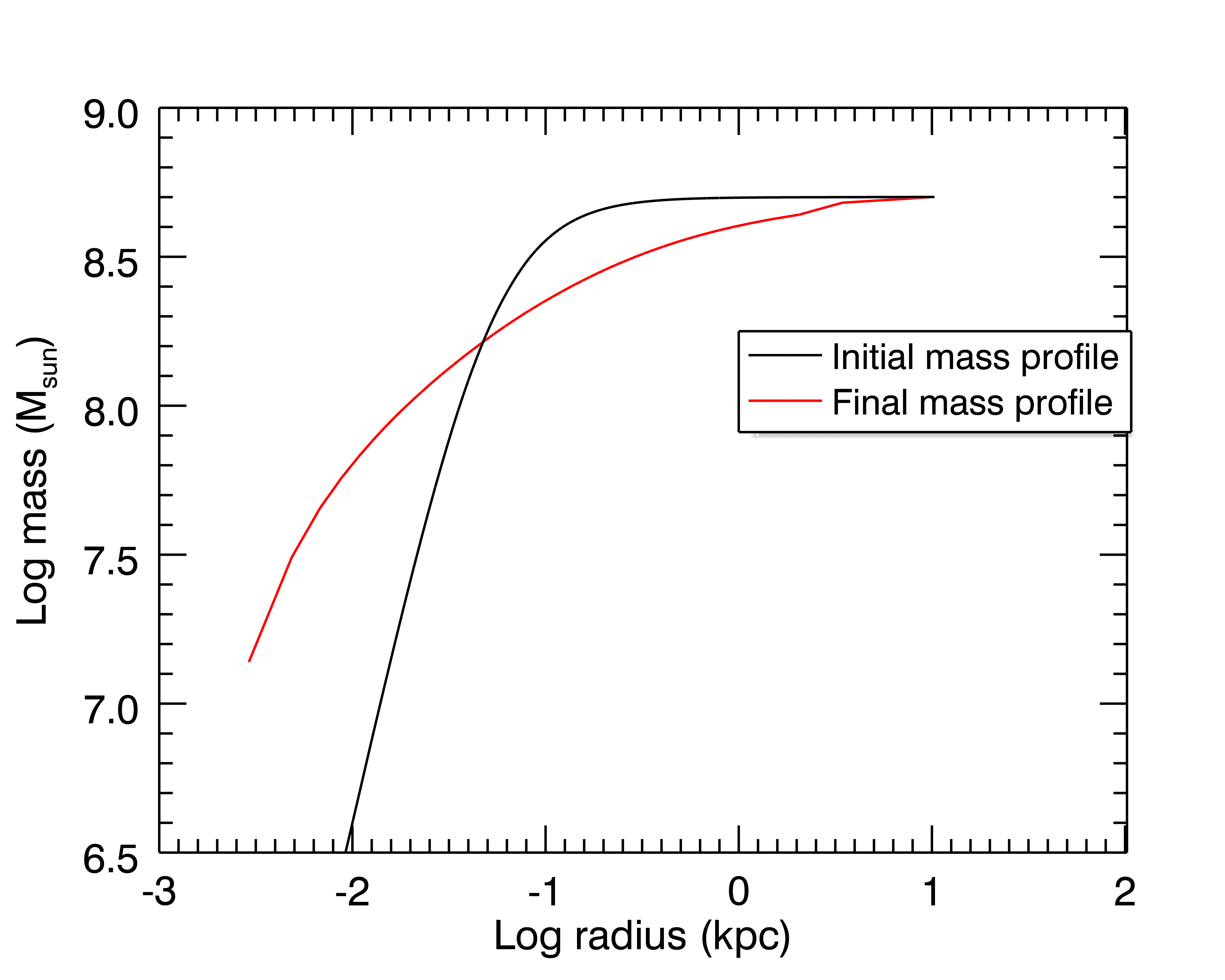}
\caption{\label{fig:corehalo} The development of the nucleus/envelope arrangement caused by the perturbation after 30 cycles. Note that the kink at The reason that the nucleus/envelope arrangement is produced is related to the aforementioned preferential selection of radial orbits. The cause will be examined in detail in section \ref{sec:bimodal}}.
\end{figure}

\subsection{Path to convergence}

Since we now have a robust grasp of the phenomenology of the attractor the final step before attempting to understand the mechanisms at work is to summarise the convergence towards to the attractor by examining the flow of individual bins. First we see if the similarities in large scale behaviour between different simulations are mirrored in similarities at the scale of a single bin. For this section we use only those simulations carried out in Newtonian gravity. Given that the binning used by our analysis codes is radius-based not mass-based the necessary differences in density profile due to differing gravitational paradigms would cause a systematic error in any side-by-side comparison.

\begin{figure}
\includegraphics[width=84mm]{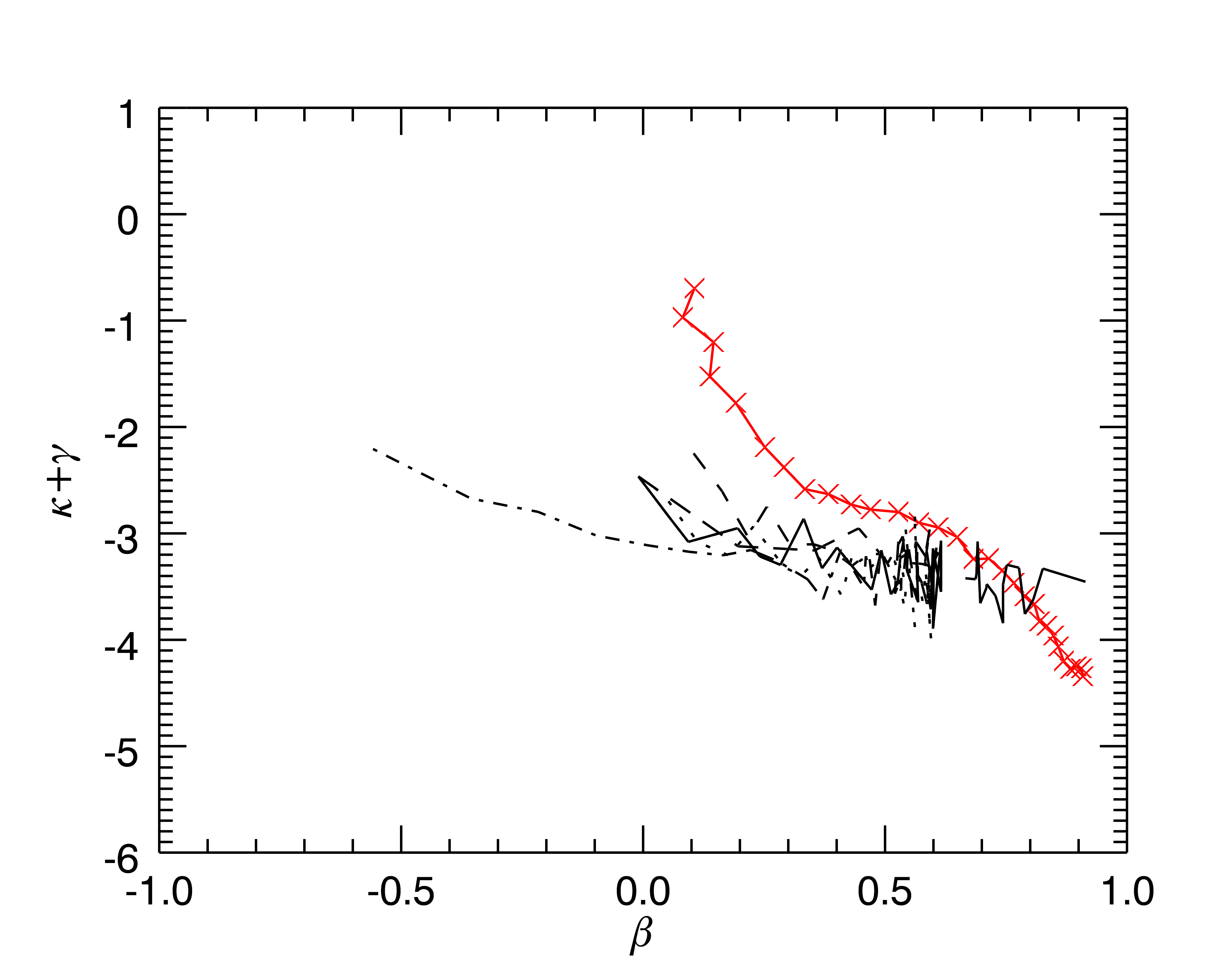}
\caption{\label{fig:bin20} The position of the $20^{th}$ bin from the 6 Newtonian simulations thus far. It is not important which line represents which simulation as all that the plot needs to convey is that the bins all move along the same path.}
\end{figure}

As we see in Figure \ref{fig:bin20}, the $20^{th}$ radial bin from the chosen simulations all follow a very similar path over the course of their evolution towards the attractor. That we have chosen the $20^{th}$ bin for this is arbitrary and largely for aesthetic reasons as there are 80 similar plots for the other bins all of which show similar agreements. The poorest agreement is found in the furthest bins as the low number of particles at those radii makes them more susceptible to statistics while the agreement of the innermost bins is harder to demonstrate visually as they move only a small distance.

With this information we can now construct a plot that shows how any radial bin should move if dropped into a simulation anywhere in our parameter space. We take all the bins in all simulations and put them in the same space before using a 2D histogram to bin that space. We create a simple `velocity' of each bin according to it's change in position between one step and the next before placing those vectors in our histogram bins and averaging them together. We can now drop test particles in our `bin velocity field' and see how they move.

\begin{figure}
\includegraphics[width=84mm]{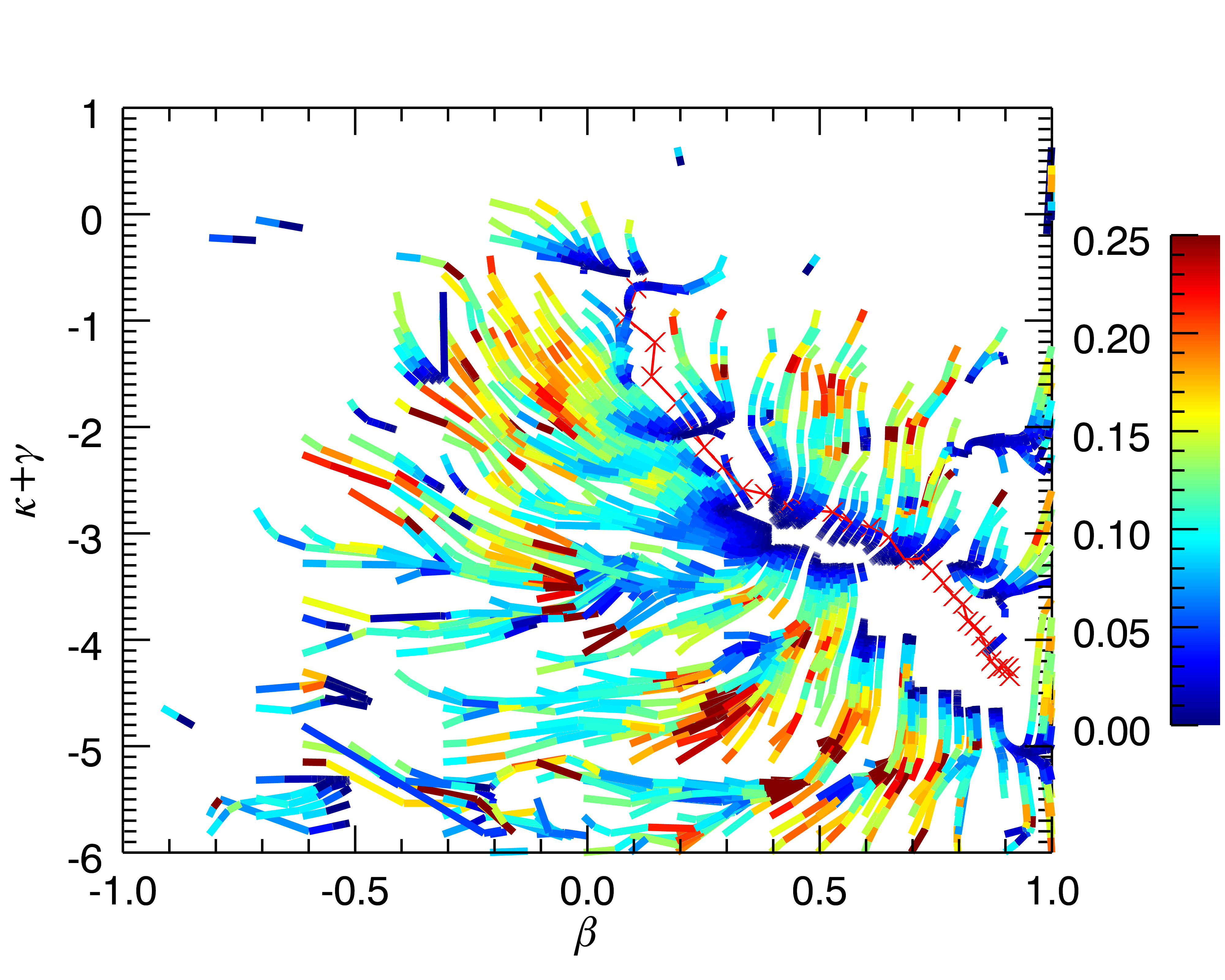}
\caption{\label{fig:stream} A representation of the attractor in terms of a vector field. Hotter colours represent larger rates of change of beta, $\frac{\Delta\beta}{\Delta t}$, in the parameter space between perturbations.}
\end{figure}

Figure \ref{fig:stream} can be interpreted as showing the vector field that the attractor creates in the parameter space and thus shows the paths that bins will move down if they are placed in a simulation. The plot nicely summarises both the apparent universality of the attractor and also allows for prediction of the state of intermediate, partially converged systems. Note that the need for discretised particles and a lack of data about certain tangential anisotropies (as the simulations evolved through the space too fast) has created some blank areas around $\beta=-0.5, \gamma+\kappa=-5$. This is unavoidable and is due to the compromise between a scarcity of very tangentially anisotropic data points and a very large number of radially anisotropic points.

\subsection{Exploration of a possible mechanism - bimodal perturbation}
\label{sec:bimodal}
So far we have several clues as to what causes the convergent behaviour. We know that the characteristics of the initial system and the environment are irrelevant and that altering the constants used in the perturbation only changes the intermediate steps and not the final conclusion. Using the fact that, as seen in section \ref{sec:flowtime}, changing the amount of time for the flow phase does not have any impact on the rate of convergence we can suppose that maybe the flow phase only acts to allow the system to find the nearest Jeans'-stable state and does not actually contribute to the attractor itself beyond the small changes seen in section \ref{sec:detail}. We might then ask if the attractor is caused primarily by the repeated action of the perturbation scheme and that the attractor represents the closest Jeans'-stable state to some limiting point for the perturbation scheme.

One potential insight into this comes from the realisation that all of our algorithms are actually special cases of a broader class of perturbation that always gives rise to the same behaviour for well-understood reasons. First, we will run down the various permutations of the algorithm and show why they all accomplish the same thing and can be thought of as being related. After establishing that one particular algorithm can be used to represent all of the other variants tested so far we will show why we would \emph{expect} that algorithm to lead to some convergent state.

As we have previously discussed, the decision to implement a perturbation that applies different factors on each velocity component has the principle effect of slowing down the convergence as the shock is not as strong as that provided by a similar scheme that uses the same factor for all three. Apart from this difference in convergence rate, however, the algorithms have the same effect and underlying principle and thus we can think of the former being a special case of the latter.

With that said, imagine a fairly weak perturbation (of the latter variety) where $0.75<f<1.25$ and consider the effect of applying this perturbation $n$ times in succession whilst allowing no opportunity for relaxation in between. We can see that, statistically, a small population will feel a perturbation like $0.75^n<f<1.25^n$ which will correspond to a more pronounced perturbation. We then posit a limiting case for these perturbations where $n$ cycles of any given perturbation will have the impact of a single perturbation where $0<f<2$, the strongest perturbation that we have employed so far and strongest possible symmetric perturbation.

So, all perturbations of the form $1-C<f<1+C$ can be thought of as reducing to $0<f<2$ over a sufficient number of iterations. Now we apply the same line of thinking; what do multiple applications of this algorithm reduce to? In this instance, a small population of particles will end up with velocities of almost zero while a small population will experience a significant boost. This time, when we apply the algorithm multiple times we will not be significantly change the energy in the population at $v\sim0$ as the most we can do is double the particles' velocity but at worst we can set it to exactly zero. This means that every time we perform the perturbation we increase the size of the low speed population while, in order to preserve energy conservation, we pump more and more energy into a small population of escapers.

The conclusion of this line of reasoning is that, in an extreme, theoretical limit, all of the perturbation schemes that perturb by this kind of random scaling cause a large population go undergo radial-infall as their energy is slowly sapped away and fed into a smaller population of particles on high energy orbits that form an envelope. This interpretation is supported by previous evidence showing an increase in central density even as the outer edges of the bound system move further and further out. To test this in the most violent limit we implement an algorithm that implements a bimodal scaling rather than a uniform one; particles are either put into radial infall by having their speed set to almost zero or are boosted into the high energy envelope. An example of such a perturbation could be that 90\% of the particles are induced to radial-infall but the 10\% envelope has $10$ times it's original energy in order to make up conservation.

For our tests an algorithm was designed that either induced radial infall or set the particle traveling at it's escape speed in such a way that, in the statistical limit, energy conservation is implied. So that the radial infall is not too perfect and to allow for the effects of the pre-existing velocity dispersion to have an effect we set the lower limit to be just more than zero; $s=0.01$.

\begin{equation}
\text{where }f=\frac{1-\frac{v_{old}^2}{v_{esc}^2}}{1-s\frac{v_{old}^2}{v_{esc}^2}}
\end{equation}
\begin{equation}
v_{old}\rightarrow v_{old}\times\left\{
  \begin{array}{l l}
    \sqrt{\frac{1-f \times s}{1-f}} & \quad \text{if $rand_u[0,1] > f$}\\
    s & \quad \text{if $rand_u[0,1] < f$}\\
  \end{array} \right.
\end{equation}

\begin{figure}
\includegraphics[width=84mm]{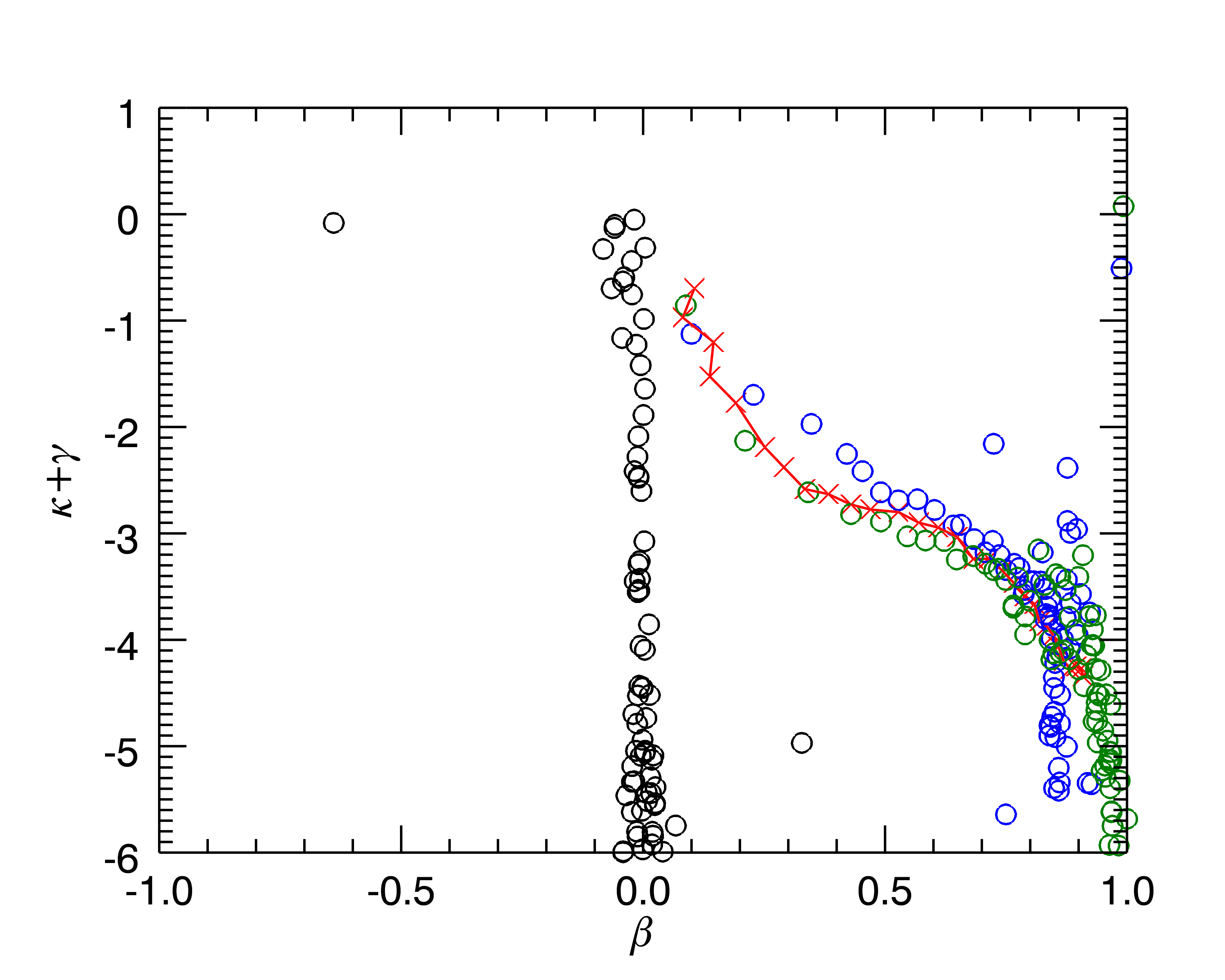}
\caption{\label{fig:bimodal} The almost immediate convergence using the bimodal algorithm showing the initial isotropic system (black), the system after only \emph{one} perturbation (blue) and after two perturbations (green)}
\end{figure}

What we see in figure \ref{fig:bimodal} is striking. After only one perturbation the system is almost exactly on top of the attractor and only one perturbation later the system has completely converged. It is interesting that the system appears to `overshoot' the attractor after the first perturbation and only settle onto it after the second kick. This is possibly because the initial perturbation is so unrealistically violent that it simply takes longer than the allowed $3\ T_{dyn}$ to equilibriate.

This suggests that the prior line of reason is valid in so far as all perturbation algorithms that we have dealt with so far are actually special, slower cases of the bimodal scheme and that the attractor can be generated simply by forcing the system to undergo our radial-infall-plus-ejection scheme and then letting the system become Jeans' stable again.

This gives us a natural explanation for why the initial anisotropy profile and theory of gravity do not affect the attractor under such perturbation schemes. Initial anisotropy does not matter as the perturbation loses that information with every successive perturbation by, in the above theoretical limit, putting all the particles on highly radial orbits by either catapulting them out to the fringes of the system or by instantaneously stopping all motion. Gravity makes no difference as the attractor here is formed by radial infall which is not a behaviour who's existence is dependent on the specifics of the theory of gravity beyond the details of the speeds involved.

This also explains interesting results from HJS where a perturbation that acts only on the radial velocity components will not lead to convergence whereas one that acts only on the tangential components will. In the tangential-only case, we can take the limiting cases where the particle ends up with no angular momentum and ends up on a purely radial orbit or is boosted onto a highly elliptical orbit. This picture is very similar to the infall mechanism and thus the same outcome is to be expected. However, if only the radial components are changed then the angular momentum of the particle cannot be removed. This means we have the same high energy limiting case but now the low energy limiting case is a circular orbit. This leads to a very radial outer envelope, as before, but a distinct tangential anisotropy in the nucleus regions that can be seen as a distinct `S-shape' bend in the anisotropy curve.

\begin{figure}
\includegraphics[width=84mm]{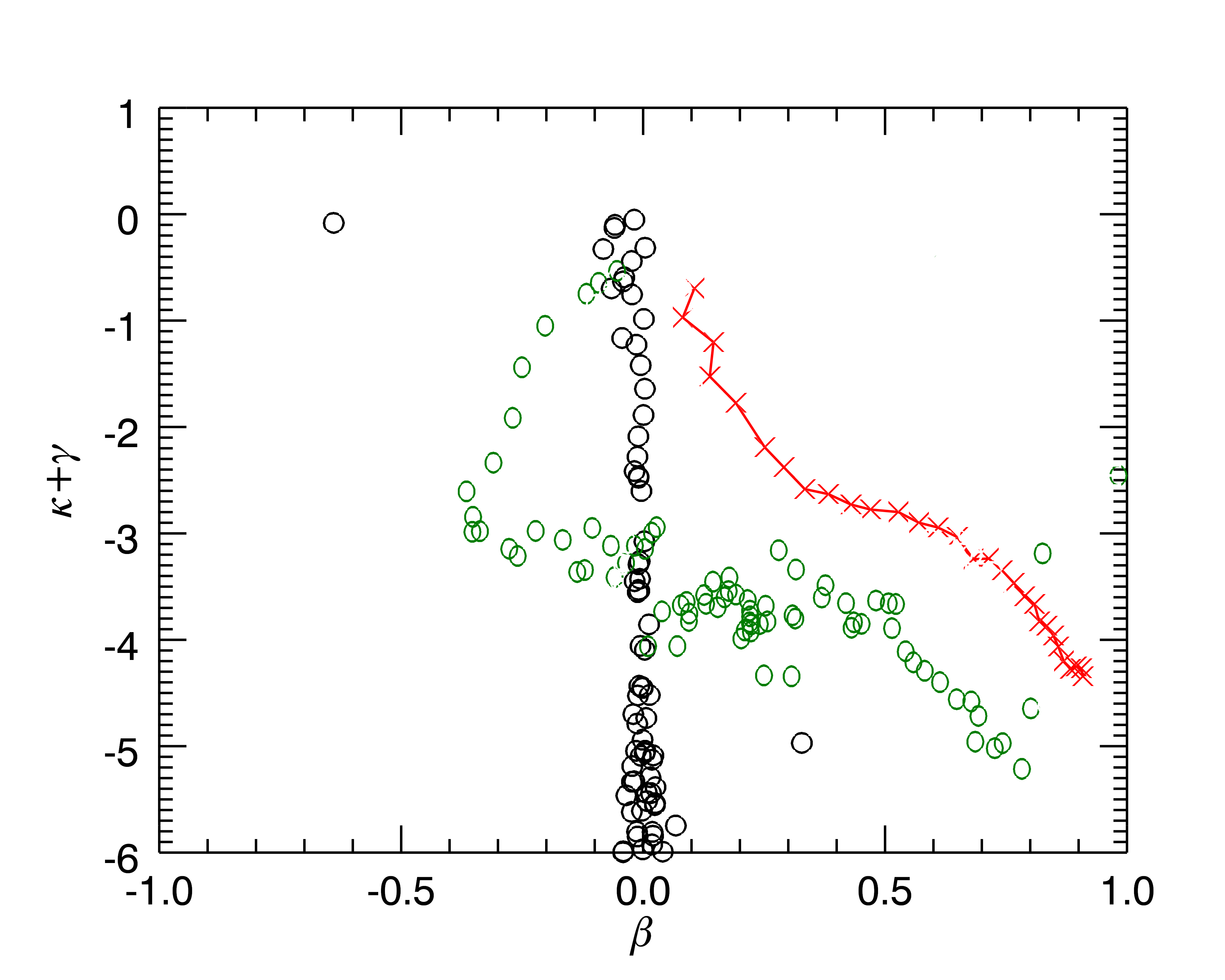}
\caption{\label{fig:radial} Altered shape of the attractor due to a strong radial bias in the perturbation algorithm.}
\end{figure}

This kind of behaviour can emerge in any code with a strong radial bias in the perturbation. The results in figure \ref{fig:radial} come from runs which attempted to conserve both angular momentum and kinetic energy. After conserving angular momentum the only way to conserve energy was to only alter the radial velocity components of each particle so as not to disturb the momentum as simple iteration towards some unrestricted and fully self-consistent solution could not guarantee convergence and, if it did, would have lost a large amount of information about the form of the intended perturbation.

\subsection{Connection to Radial Orbit Instability (ROI)}

In the previous section we have shown that the perturbation schemes being used in our work all appear to reduce the problem to one of the evolution of a system under ROI. Much work has already been done in studying whether ROI is the explanation for the universality of the NFW profile in dark matter halos \citep{Macmillan2006,Bellovary2008,Lapi2011} and several universal profiles and convergent behaviours have emerged. While these results are not exclusive to ROI they are all related to the mechanisms of collapse and, given the discussion of the previous section, a very radial mode of collapse, such as ROI, would seem to be the most promising lens through which to view our results.

Before continuing it is important to point out that the mechanism we use is not pure ROI as it forces the system to undergo an ROI-like collapse regardless of the initial susceptibility of the system to ROI. For example, simulations in \citet{Bellovary2008} that used isotropic initial conditions found them to be stable against ROI compared to radially anisotropic ones and \citet{Barnes2009} found a strong tendency towards the development of triaxiality. Our simulations evolve regardless of initial anisotropy and section \ref{sec:sphersym} showed our simulations remain generally spherical as our perturbation has no preferred axis. The triaxiality in the deep MOND simulations is interesting but ultimately a statistical effect of having a much more extended system to start with. Accordingly, the following are currently presented as interesting connections between our work and ROI with the strength of the connection a topic of ongoing investigation.

\subsubsection{Universal beta profile}
In many simulations it was observed that the anisotropy profile would tend towards a common profile as well, although more complex than a clean power-law \citep{Bellovary2008,Lapi2011}, with an isotropic nucleus that rises smoothly to radial anisotropy in the outer edges. There is some tolerance within this, however, as \citet{Bellovary2008} found that the initial anisotropy profile did leave a lingering impression on the final profile

\begin{figure}
\includegraphics[width=84mm]{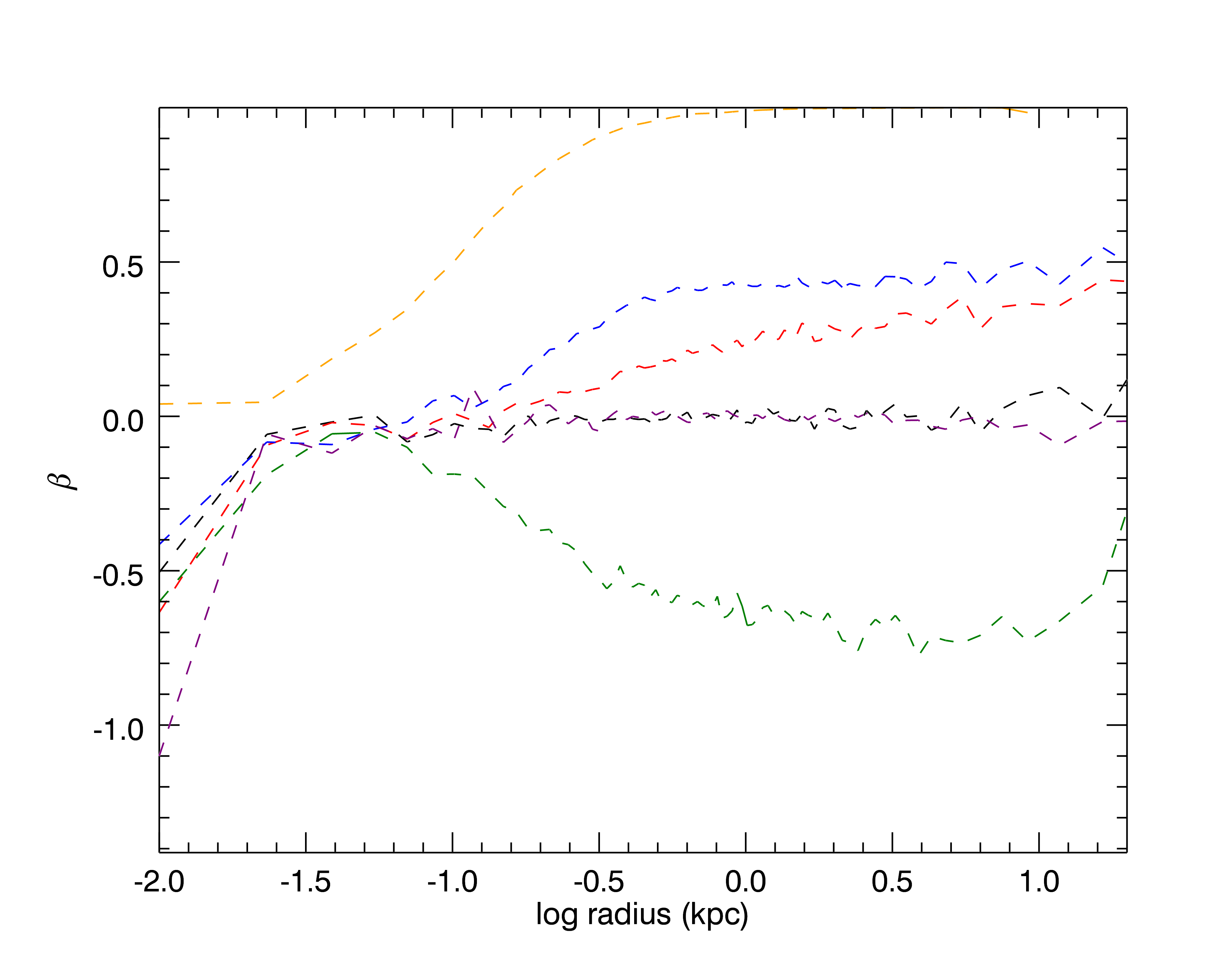}
\includegraphics[width=84mm]{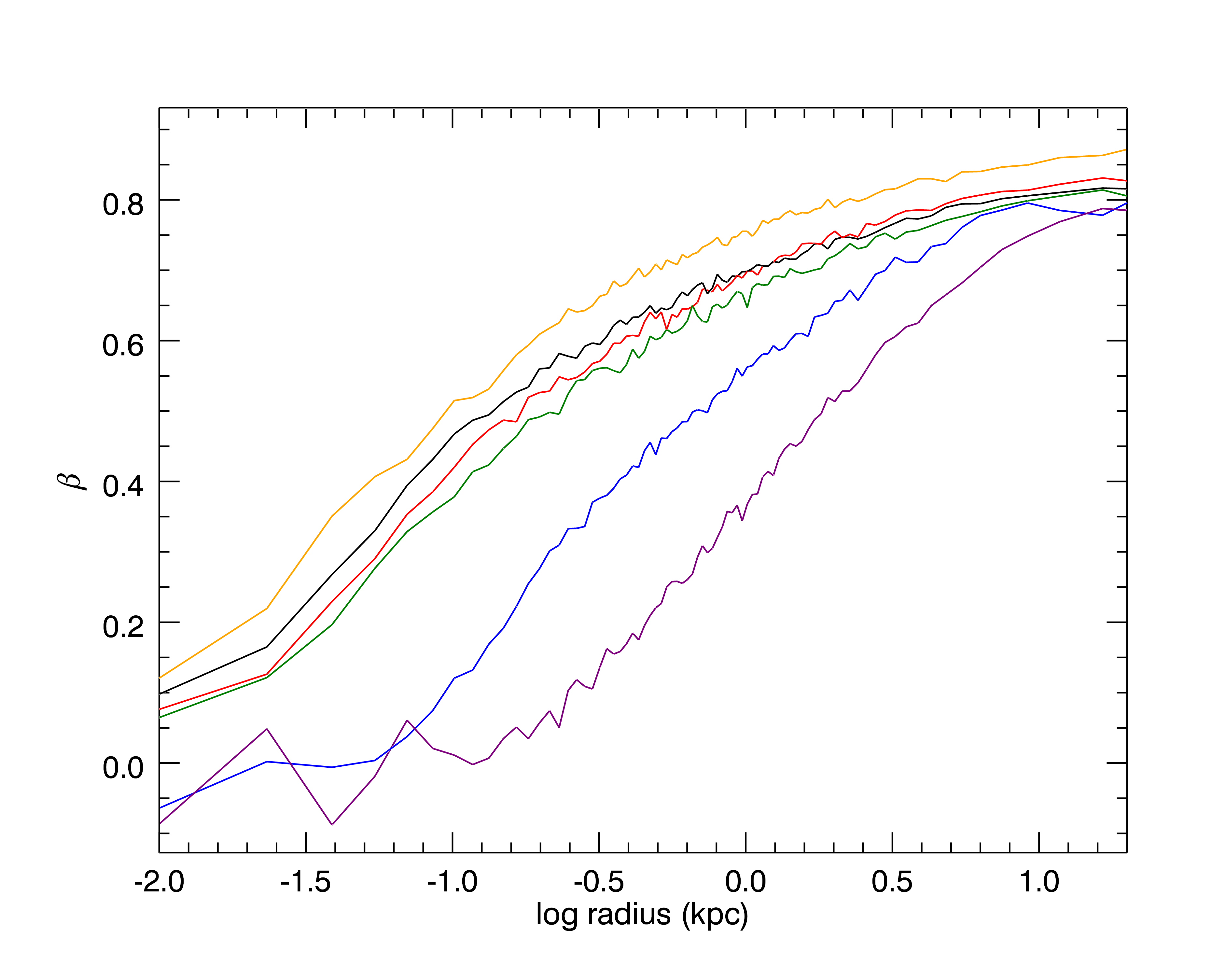}
\caption{\label{fig:betaprofile} Evolution of the anisotropy profile in the Newtonian isotropic (black), radial (blue), tangential (green) and extremely radial (orange) models as well as the MONDian isotropic (purple) and radial (blue) models from their initial states (dotted lines/top panel) to their overall final states (solid lines/bottom panel).}
\end{figure}

We so find good convergence among the Newtonian simulations although again we find that the MOND simulations stand out somewhat in figure \ref{fig:betaprofile} as well as the extremely radial simulations. We would expect this kind of convergence regardless of any discussion about ROI as we have already discussed in detail the physics that leads to these profiles.

\subsubsection{Convergence of $\rho$/$\sigma^3$}
One important result, although not one exclusive to ROI, was that systems formed by mergers (such as DM halos) display convergence in a quantity that proxies phase-space density \citep{Taylor2001,Dehnen2005}:

\begin{equation}
\label{eqn:powerlaw}
\frac{\rho}{\sigma_r^3}\propto r^{-1.9}
\end{equation}

\begin{figure}
\includegraphics[width=84mm]{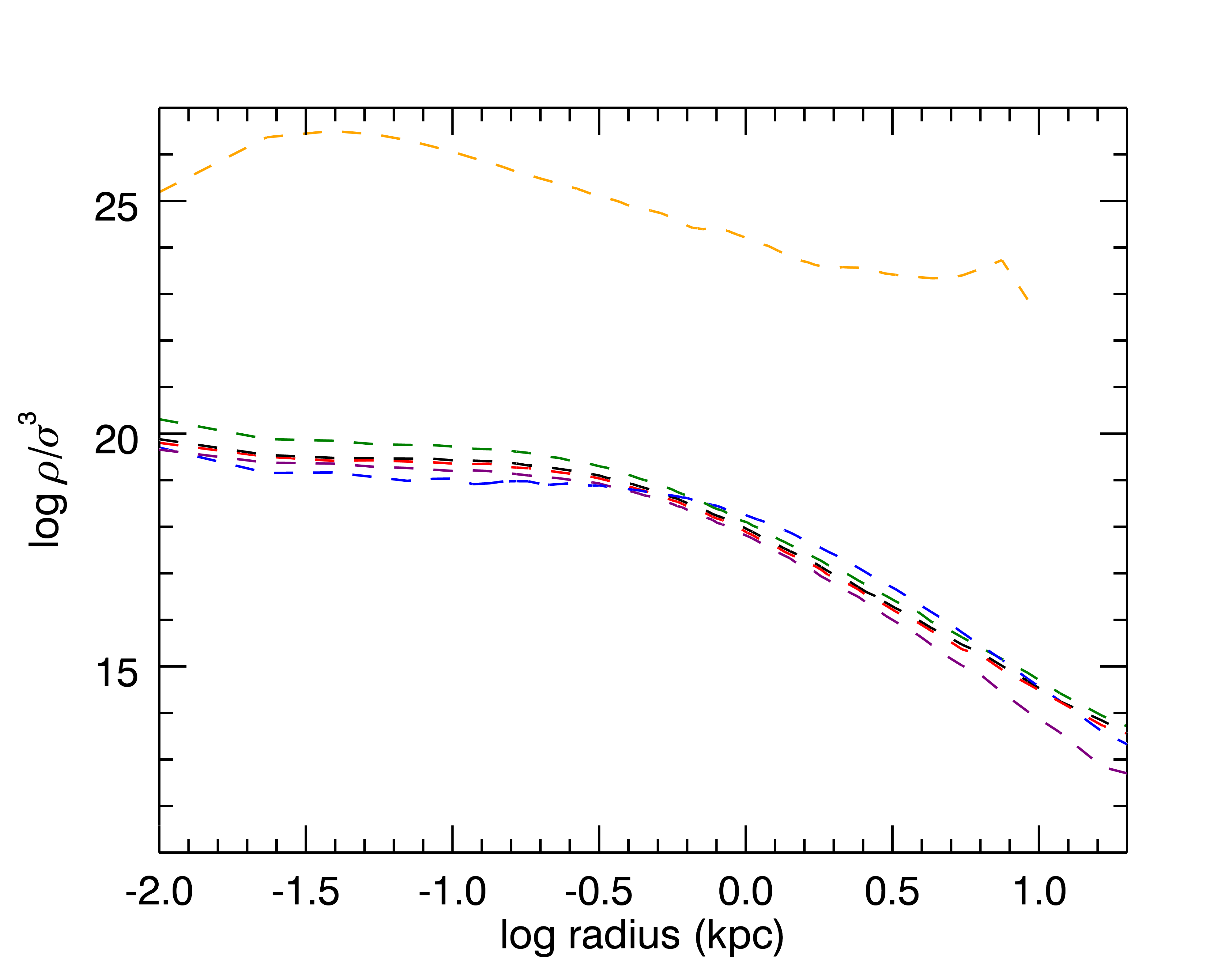}
\includegraphics[width=84mm]{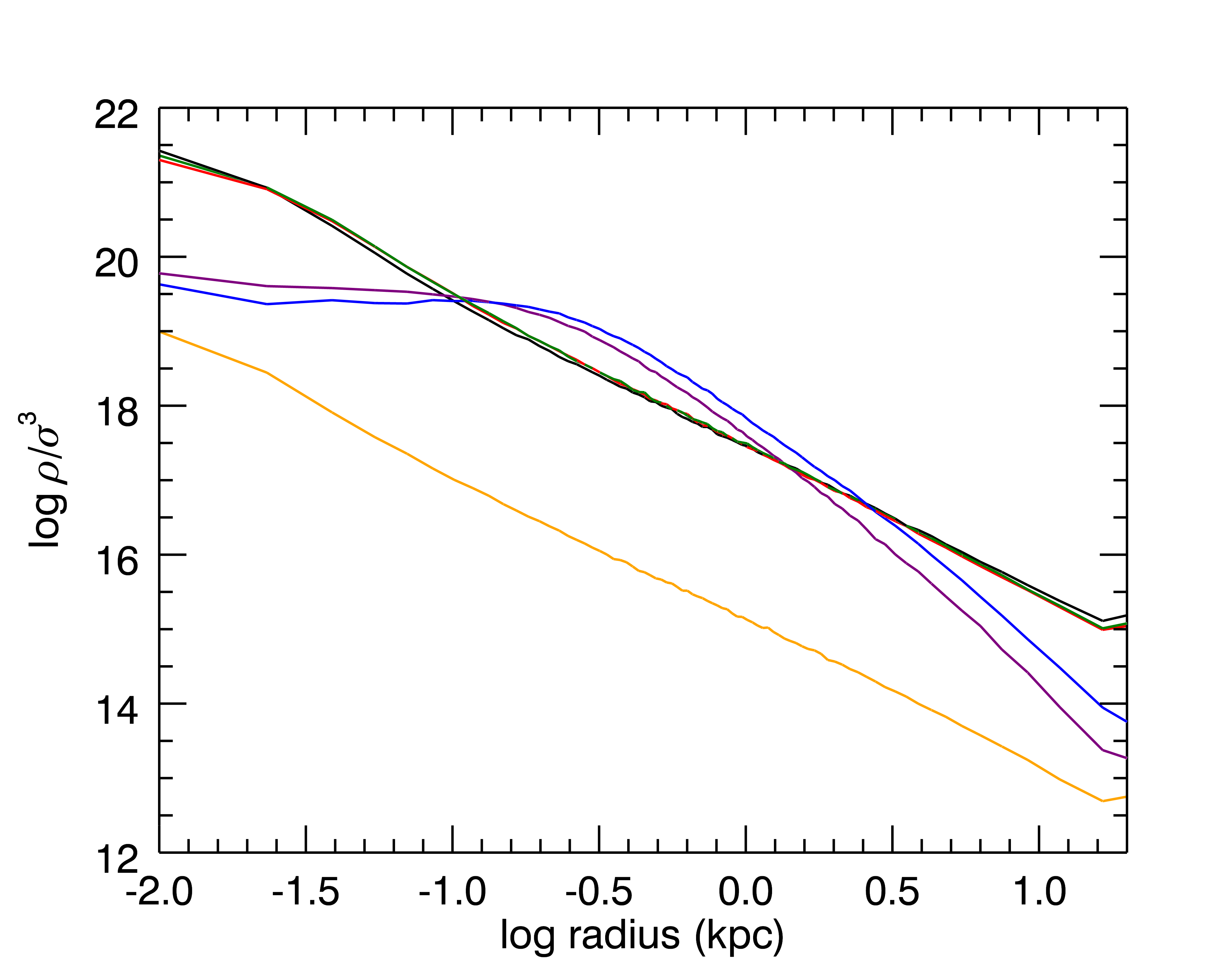}
\caption{\label{fig:powerlaw} Evolution of phase-space density in the Newtonian isotropic (black), radial (blue), tangential (green) and extremely radial (orange) models as well as the MONDian isotropic (purple) and radial (blue) models from their initial states (dotted lines/top panel) to their overall final states (solid lines/bottom panel).}
\end{figure}

As noted in the references, the peculiar thing about this convergence is that neither $\rho$ nor $\sigma_r^3$ are themselves convergent, to a power-law or otherwise. However, the significance of this quantity does suggest that there may be some connection between it and our $\gamma+\kappa$ axis. Here, in figure \ref{fig:powerlaw}, all we investigate is whether our simulations develop the same power-law during convergence.

Figure \ref{fig:powerlaw} shows us firstly that all the models start of off with very similar phase-space density profiles with the exception of the extremely radial case. However, \emph{all} the Newtonian models, after evolution, tend towards the $r^{-2}$ (equation \ref{eqn:powerlaw}) power-law reported by \citet{Taylor2001} and \citet{Dehnen2005} and in fact end up with profiles that are closer together than their already good agreement to begin with. Standing out from this close convergence is the extremely radial case which ends up with a translated $r^{-2}$ power-law. Also interesting is that this tight convergence is completely absent from the MONDian profiles which retain good agreement with each other but have barely evolved from their initial conditions.

What this shows is that our perturbation, developed from HJS's attempt to simply model halo mergers, does seem to display some characteristics of a system that has undergone repeated mergers. Accordingly, we may use this power-law as a possible avenue of investigation into our attractor formula. Particularly in light of the notable connections to the next result.

\subsubsection{$\beta$ - $\gamma$ relation}
Seeing as though we now have two convergent properties, both involving velocity dispersions and density, it is not surprising that there is a relation between both $\beta$ and $\gamma$, the density slope, that is also a convergent result. Many studies have found evidence for a relationship between the two \citep{Huss1999,Barnes2005,HansenMoore2006,Hansen2006,Macmillan2006,Bellovary2008,Lapi2011} and \citet{HansenMoore2006} went as far as positing a relationship of $\beta(r)=-0.15+0.2\gamma$. Although this result was not linked directly to ROI, instead being found in every relaxed structure, it should still apply and several papers have used it for such.

So, as we can see there are already several `attractor' solutions relating to $\gamma$ and $\beta$ in various parameter spaces if a system undergoes ROI. With this in mind, it is perhaps not surprising that convergent results are also a feature of our perturbations. We cannot draw too many conclusions as we know that our perturbations are not producing textbook ROI and, interestingly, our MONDian simulations seem to produce the attractor despite not adhering to various convergent behaviours normally seen in ROI. However, the amount of established relationships in ROI systems does lead towards the possibility of finding an analytical solution to the attractor from first-principles, at least in Newtonian gravity, as well as offering a clear set of criteria to test for the emergence of the attractor.

\subsection{Ongoing tests}

Since we find that all the tests so far are related to this radial infall mechanism to some degree we are designing tests that specifically exclude radial infall scenarios by not using randomised rescaling of velocity components. Work is ongoing into a perturbation scheme where a velocity distribution function is chosen \emph{a priori} and the system is forced to follow that curve. It meets the requirements of a perturbation in this context as it moves particles in velocity space into a non-equilibrium solution however, unlike previous algorithms, information about anisotropy should be preserved as only the total speeds are being effected and the predetermined distribution function should prevent the formation of the `nucleus' and `envelope' populations that drive the attractor in all previous examples.

Care must be taken to prescribe a distribution function with well understood behaviour otherwise the behaviour of the attractor could be confused with eccentric behaviour caused by choosing an unphysical distribution function.

Investigations are ongoing as to whether the results from ROI work can be used to predict an analytic solution to the attractor by narrowing down the relationship between our perturbation simulations, MOND and the more general body of work surrounding ROI phenomenology. While our systems are not exactly inducing ROI, there appear to be a large enough number of similarities for ROI to be a worthwhile avenue of inquiry.

A different set of perturbations are considered in Sparre \& Hansen (2012, \emph{[in prep]}), where the importance of violent relaxation is discussed.

\section{Conclusions}

We have tested a variety of initial conditions, gravity models and perturbation schemes and found that in each case the system will tend towards a single set of solutions as described by HJS. The attractor appears independent of initial anisotropy and prevailing theory of gravity due to the perturbations tested thus far all creating a system heavily influenced by radial-infall behaviours. This appears to be the only identifiable driving mechanism at this time as virialisation, Jeans' stability and the Antonov stability laws do not suggest any pressure to find new equilibria.

As noted by HJS it is possible to create algorithms that defy the attractor such as the radial-only algorithms that have been mentioned previously. That particular case is not a cause for concern as it is clearly a very unrealistic perturbation method. Work is ongoing into finding a realistic algorithm that, according to the infall approach, should not lead to convergence and to that end an algorithm based on prescribed velocity distribution functions is being designed.

Finally, it is interesting to look at our density profiles compared against NFW profiles. This is purely a curiosity as, while it would be very interesting to find that the attractor is also an attractor in density space, there is no reason to expect it to be. First, it is important to note that, due to the different shape of the potential-density pair for MOND, the MONDian Plummer spheres have the same shape but a lower central density so that more matter can be present in the outer regions.

\begin{figure}
\includegraphics[width=84mm]{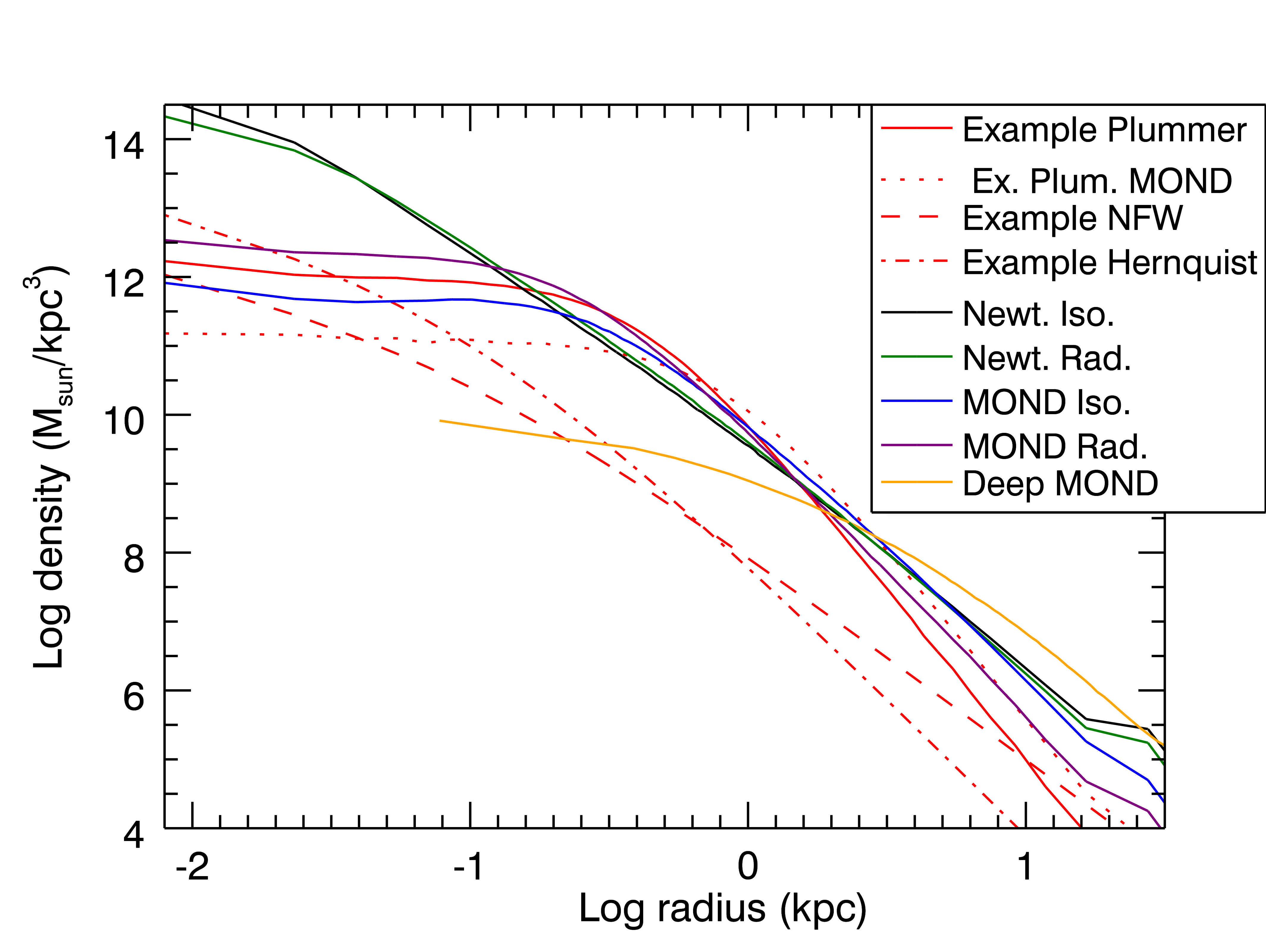}
\caption{\label{fig:finalden} Final density profiles from a selection of simulations contrasted with an arbitrary NFW profile, arbitrary Hernquist profile and the initial conditions of the Newtonian and weak MOND Plummer profiles.}
\end{figure}

From figure \ref{fig:finalden} we see that the models have changed noticeably from their initial configurations. We see that the attractor does manifest some effect as the `nucleus/envelope' arrangement is visible here as the increased density in the very centre and the outer regions has the effect of producing an almost Hernquist-like profile. The significantly steeper and higher central profiles for the Newtonian models compared to MOND are due to the higher core density of such models as, for instance, the very extended deep MOND profile has an apparently very small core density for much the same reason. The difference is due to the perturbation increasing the core density relative to it's initial state which is proportionally smaller in models with more pronounced MONDian effects. It is unlikely that this represents a statement about the behaviour of particles in a MONDian regime.

Overall, this is an interesting and robust phenomena that we will continue to study. In the event that subsequent work demonstrates that the radial infall mechanism is only a small effect relevant to this class of perturbation then the consequences and opportunities for observational prediction are significant.

\section{Acknowledgments}

We gratefully acknowledge the help and advice of Martin Sparre and Diana Juncher at the Niels Bohr Insitute throughout this work. We would also like to thank the Niels Bohr Institute's Dark Cosmology Centre for generously funding two weeks for the St Andrews group to come to Denmark for detailed discussion. The Dark Cosmology Centre is funded by the Danish National Research Foundation. XW thanks Ortwin Gerhard and Flavio de Lorenzi for sharing the code of generating anisotropic models by using circularity functions and Lucy's method. XW acknowledges the postdoc fellowship from the Dynamics Group in Max-Planck-Institut fuer Extraterrestrische Physik.

\bibliographystyle{mn2e}
\bibliography{attractor_paper}

\begin{thebibliography}{}

\bibitem[\protect\citeauthoryear{Barnes, Lanzel \& Williams}{Barnes
  et~al.}{2009}]{Barnes2009}
Barnes E.,  Lanzel P.,    Williams L.,  2009, The Astrophysical Journal, 704,
  372

\bibitem[\protect\citeauthoryear{Barnes, Williams, Babul \& Dalcanton}{Barnes
  et~al.}{2005}]{Barnes2005}
Barnes E.,  Williams L.,  Babul A.,    Dalcanton J.,  2005, The Astrophysical
  Journal, 634, 775

\bibitem[\protect\citeauthoryear{Bekenstein \& Milgrom}{Bekenstein \&
  Milgrom}{1984}]{Bekenstein1984}
Bekenstein J.,  Milgrom M.,  1984, The Astrophysical Journal, 286, 7

\bibitem[\protect\citeauthoryear{Bellovary, Dalcanton, Babul, Quinn, Maas,
  Austin, Williams \& Barnes}{Bellovary et~al.}{2008}]{Bellovary2008}
Bellovary J.,  Dalcanton J.,  Babul A.,  Quinn T.,  Maas R.,  Austin C.,
  Williams L.,    Barnes E.,  2008, The Astrophysical Journal, 685, 739

\bibitem[\protect\citeauthoryear{Ciotti, Londrillo \& Nipoti}{Ciotti
  et~al.}{2006}]{Ciotti2006}
Ciotti L.,  Londrillo P.,    Nipoti C.,  2006, The Astrophysical Journal, 640,
  741

\bibitem[\protect\citeauthoryear{Dehnen \& McLaughlin}{Dehnen \&
  McLaughlin}{2005}]{Dehnen2005}
Dehnen W.,  McLaughlin D.,  2005, Monthly Notices of the Royal Astronomical
  Society, 363, 1057

\bibitem[\protect\citeauthoryear{Dubinski \& Carlberg}{Dubinski \&
  Carlberg}{1991}]{Dubinski1991}
Dubinski J.,  Carlberg R.,  1991, The Astrophysical Journal, 378, 496

\bibitem[\protect\citeauthoryear{Gerhard}{Gerhard}{1991}]{Gerhard1991}
Gerhard O.,  1991, Monthly Notices of the Royal Astronomical Society, 250, 812

\bibitem[\protect\citeauthoryear{Hansen, Juncher \& Sparre}{Hansen
  et~al.}{2010}]{Hansen2010}
Hansen S.,  Juncher D.,    Sparre M.,  2010, The Astrophysical Journal, 718,
  L68

\bibitem[\protect\citeauthoryear{Hansen \& Moore}{Hansen \&
  Moore}{2006}]{HansenMoore2006}
Hansen S.,  Moore B.,  2006, New Astronomy, 11, 333

\bibitem[\protect\citeauthoryear{Hansen, Moore, Zemp \& Stadel}{Hansen
  et~al.}{2006}]{Hansen2006}
Hansen S.,  Moore B.,  Zemp M.,    Stadel J.,  2006, Journal of Cosmology and
  Astroparticle Physics, p.~14

\bibitem[\protect\citeauthoryear{Host \& Hansen}{Host \&
  Hansen}{2011}]{Host2011}
Host O.,  Hansen S.,  2011, arXiv:0907.1097v2

\bibitem[\protect\citeauthoryear{Huss, Jain \& Steinmetz}{Huss
  et~al.}{1999}]{Huss1999}
Huss A.,  Jain B.,    Steinmetz M.,  1999, The Astrophysical Journal, 517, 64

\bibitem[\protect\citeauthoryear{Jeans}{Jeans}{1915}]{Jeans1915}
Jeans J.,  1915, Monthly Notices of the Royal Astronomical Society, 76, 70

\bibitem[\protect\citeauthoryear{Jing}{Jing}{2000}]{Jing2000}
Jing Y.,  2000, The Astrophysical Journal, 535, 30

\bibitem[\protect\citeauthoryear{Lapi \& Cavaliere}{Lapi \&
  Cavaliere}{2011}]{Lapi2011}
Lapi A.,  Cavaliere A.,  2011, The Astrophysical Journal, 743, 127

\bibitem[\protect\citeauthoryear{Londrillo \& Nipoti}{Londrillo \&
  Nipoti}{2008}]{NMODY}
Londrillo P.,  Nipoti C.,  2008

\bibitem[\protect\citeauthoryear{McGaugh \& De~Blok}{McGaugh \&
  De~Blok}{1998}]{McGaugh1998}
McGaugh S.,  De~Blok W.,  1998, The Astrophysical Journal, 499, 41

\bibitem[\protect\citeauthoryear{Macmillan J.D.and~Widrow \&
  Henriksen}{Macmillan \& Henriksen}{2006}]{Macmillan2006}
Macmillan J.D.and~Widrow L.,  Henriksen R.,  2006, The Astrophysical Journal,
  653, 43

\bibitem[\protect\citeauthoryear{Makino, Sasaki \& Suto}{Makino
  et~al.}{1998}]{Makino1998}
Makino N.,  Sasaki S.,    Suto Y.,  1998, The Astrophysical Journal, 497, 555

\bibitem[\protect\citeauthoryear{Navarro, Frenk \& White}{Navarro
  et~al.}{1996}]{Navarro1996}
Navarro J.,  Frenk C.,    White S.,  1996, The Astrophysical Journal, p.~255

\bibitem[\protect\citeauthoryear{Salucci, Lapi, Tonini, Gentile, Yegorova \&
  Klein}{Salucci et~al.}{2007}]{Salucci2007}
Salucci P.,  Lapi A.,  Tonini C.,  Gentile G.,  Yegorova I.,    Klein U.,
  2007, Monthly Notices of the Royal Astronomical Society, 378, 41

\bibitem[\protect\citeauthoryear{Taylor \& Navarro}{Taylor \&
  Navarro}{2001}]{Taylor2001}
Taylor J.,  Navarro J.,  2001, The Astrophysical Journal, 563, 483

\end{thebibliography}

\appendix

\section{Details of MONDian dynamics}
\label{sec:MOND}
For a spherical system the equation that NMODY uses \citep{NMODY} to solve for gravitational attraction is \citep{Bekenstein1984}:

\begin{equation}
\nabla\cdot\left[\mu\left(\frac{g}{a_0}\right)\nabla\phi(r)\right]=4\pi G\rho(r)
\end{equation}

to give an acceleration field, $g$, for a potential $\phi(r)$ resulting from a density profile $\rho(r)$. The $\mu(x)$ function is what dictates the strength of the MONDian modification to the straightforward Newtonian case. There are a variety of different functions that can be used, but all use x, the ratio of the strength of the acceleration field to the MONDian threshold value of $a_0=3600(km/s)^2/kpc$, and have the same asymptotic behaviour:

\begin{equation}
\mu(x) \rightarrow \left\{
  \begin{array}{l l}
    x & \quad x \ll 1\\
    1 & \quad x \gg 1\\
  \end{array} \right.
\end{equation}

Put another way, for a given potential we can calculate the acceleration under Newtonian dynamics, $g_N$, using Poisson's equation. We can relate this to the MONDian acceleration, $g$, using $\mu(x)$. Here we use a common, simple $\mu(x)$ (noting that we are still only concerned with spherically symmetric problems):

\begin{equation}
\frac{g_N}{g}=\mu\left(\frac{g}{a_0}\right)=\mu(x)=\frac{x}{1+x}
\end{equation}

In order to generate theoretical acceleration curves, such as those seen in figure \ref{fig:mondpotential}, a similar process is carried out. For the mu function described above, the process would be:

\begin{equation}
\frac{g}{g_N}=\frac{1}{\mu(x)}\equiv\nu(y)=\nu\left(\frac{g_N}{a_0}\right)=\frac{1}{2}+\sqrt(\frac{1}{4}+\frac{a_0}{g_N})
\end{equation}

Thus, any Newtonian acceleration profile can be related to a MONDian one by $g=g_N\nu(y)$. In the case of the deep MOND systems a different $\mu(x)$ is used as the system is entirely in the MOND regime. The deep MOND theoretical acceleration profiles seen in figure \ref{fig:mondpotential} are found as follows:

\begin{equation}
\nabla\cdot\left[|\nabla\phi|\nabla\phi\right]=4\pi Ga_0\rho(r)
\end{equation}
\begin{equation}
(\nabla\phi)^2=4\pi Ga_0\frac{1}{r^2}\int^r_0\rho(r)r^2dr
\end{equation}
\begin{equation}
\nabla\phi=\sqrt\frac{GM(<r)a_0}{r^2}
\end{equation}

where, in our case, $\rho(r)$ would be the profile for a Plummer model.

\section{Methods for L conservation}
\label{sec:append}
If we want to conserve angular momentum then, for example, we can require the $Lx$, $Ly$ and $Lz$ components of each bin to be equal before and after. The change in each momentum component from before and after the perturbation is assessed and each particle given an equal share of that change, ${\delta}L[x,y,z]$, to make up. Dealing in Cartesian coordinates, each velocity component affects two momentum components which means that each velocity component has two pieces of information about how it must change to move towards conservation. The subsequent change made to the velocity component is an average of these two values. For example, for $v_x$:

\begin{equation}
\label{eqn:momcon}
v_x=v_x-\frac{1}{2}\left[\frac{{\delta}Ly|z|}{(|x|+|z|)z}+\frac{-{\delta}Lz|y|}{(|x|+|y|)y}\right]
\end{equation}

with cyclical permutations for the other components. It can be shown that this approach rapidly converges to the correct global value. A qualitative explanation is that, in every circumstance, the two velocity components furthest from their required values will move to improve the two worst angular momentum offsets.

\section{Impact of sphericity on anisotropy}
\label{sec:append2}
A simple thought experiment involves a flat disc with a population of particles at all radii in circular orbits. If we then take that system and stretch it along one axis by perfectly setting up an injection of radial velocity for every particle at the correct time we will end up with a set of elliptical orbits and a non-circular system. If we now place a circular mask over the system for a binning procedure and look at the orbits of the particles we will find a mix of radial and tangential motion. The problem is that this will now look like a circular system with a degree of anisotropy rather than like a very regular elliptical system.

\bsp

\label{lastpage}

\end{document}